\documentclass[12pt]{article}
\usepackage{amsmath,amsfonts,amssymb,amsxtra}
\usepackage{graphicx}
\input epsf

\usepackage[numbers,sort&compress]{natbib}

\makeatletter\@addtoreset{equation}{section}\makeatother

\newcommand{\preprint}[1]{\begin{table}[t]  %%
             \begin{flushright}               %%
             {#1}                             %%
             \end{flushright}                 %%
             \end{table}}                     %%
\renewcommand{\title}[1]{\vbox{\center\LARGE{#1}}\vspace{5mm}}
\renewcommand{\author}[1]{\vbox{\center#1}\vspace{5mm}}
\newcommand{\address}[1]{\vbox{\center\em#1}}
\newcommand{\email}[1]{\vbox{\center\tt#1}\vspace{5mm}}

\newcommand {\la} {\left \langle}
\newcommand {\ra} {\right \rangle}
\newcommand {\lb} {\left (}
\newcommand {\rb} {\right )}

\newcommand {\CalN} {\mathcal N}

\newcommand {\CalM} {\mathcal M}

\newcommand {\BR}   {\mathbb R}
\newcommand {\BZ}   {\mathbb Z}

\newcommand {\ve}  {\varepsilon}
\newcommand {\ep}  {\epsilon}

\newcommand{\g}{\mathfrak{g}}

\renewcommand{\Re} {\mathrm{Re}}

\newcommand {\p} {\partial}

\DeclareMathOperator{\tr} {tr}

\DeclareMathOperator{\Pexp} {Pexp}

\DeclareMathOperator{\diag}{diag}

\newcommand{\SU}{SU}
\newcommand{\U}{U}
\newcommand{\SO}{SO}

\newcommand{\so}{\mathfrak {so}}

\numberwithin{equation}{section}

\begin{document}

\unitlength = .8mm

\bibliographystyle{utphys}

\begin{titlepage}
\begin{center}
\hfill \\
\hfill \\

\preprint{ITEP-TH-45/09}

\title{The 1/2 BPS 't~Hooft loops in $\CalN=4$ SYM as instantons in 2d Yang-Mills}

\renewcommand{\thefootnote}{\fnsymbol{footnote}}
\author{Simone Giombi$^{a}$ and
Vasily Pestun$^{b,\,}$\footnotemark}
\footnotetext{On leave of absence from ITEP, 117218, Moscow, Russia}

\address{Center for the Fundamental Laws of Nature
\\Jefferson Physical Laboratory, Harvard University,\\
Cambridge, MA 02138 USA\\
}

\email{$^a$giombi@physics.harvard.edu,
$^b$pestun@physics.harvard.edu}

\end{center}

\abstract{We extend the recent conjecture on the relation between a certain 1/8 BPS subsector 
of 4d $\CalN=4$ SYM on $S^2$ and 2d Yang-Mills theory by turning on circular 1/2~BPS 't~Hooft operators linked with $S^2$. We show that localization predicts that these 't~Hooft operators and their correlation functions with Wilson operators on $S^2$ are captured by instanton contributions to the partition function of the 2d Yang-Mills theory. 
Based on this prediction, we compute explicitly correlation functions involving the 't~Hooft operator, and observe precise agreement with $S$-duality predictions.}

\vfill

\end{titlepage}

\eject

\tableofcontents

%%% Local Variables: 
%%% mode: latex
%%% TeX-master: "main"
%%% End:

\newcommand{\gym}{g_{4d}}

\section{Introduction}
The ${\cal N}=4$ Super Yang-Mills theory is believed to enjoy an exact quantum symmetry, known as $S$-duality \cite{Montonen:1977sn,Goddard:1976qe,Witten:1978mh,Osborn:1979tq}, which relates weak coupling to strong coupling physics, and can be thought of as a non-abelian generalization of the familiar electric-magnetic duality of Maxwell's theory. More precisely, the $S$-duality symmetry of ${\cal N}=4$ SYM with gauge group $G$ acts on the complex coupling $\tau = \frac{\theta}{2\pi}+\frac{4\pi i}{\gym^2}$ as 
\begin{equation}
\tau \rightarrow {^L}\tau = -\frac{1}{n_{\g} \tau}\,,
\label{S-tau-map}
\end{equation}
and exchanges the gauge group $G$ with its $S$-dual, or \emph{Langlands dual}, group ${^L}G$ \cite{Goddard:1976qe,Kapustin:2006pk,Argyres:2006qr,Witten:2009mh}. Here $\g$ denotes the Lie algebra of $G$, and $n_{\g}=1$ for the simply laced Lie algebras, $n_{\g}=2$ for $\so(2N+1)$, $\mathfrak{sp}(N)$, $\mathfrak{f}_4$ and $n_{\g}=3$ for $\mathfrak{g}_2$. The transformation (\ref{S-tau-map}), together with the elementary symmetry $\tau \rightarrow \tau+1$, generate an infinite group $\Gamma$ which is a discrete subgroup of $SL(2,\mathbb{R})$. For the simply laced Lie algebras, this is just the familiar $SL(2,\mathbb{Z})$ modular group acting on $\tau$.

Under $S$-duality, the electric and magnetic degrees of freedom are mapped into each other. In particular, the Wilson loop operator, which describes an electric charge running along a contour in space-time, should be mapped to its magnetic counterpart, the 't~Hooft loop. In a gauge theory
with a gauge group $G$, a Wilson loop operator is defined as the holonomy of the gauge field along a given contour (or, in supersymmetric theories, as a suitable generalization involving scalar fields), and hence is labeled by a choice of representation $R$ of the gauge group $G$. On the other hand, a 't~Hooft operator cannot be described as a functional of the fields, but rather is defined by requiring that in the path integral we integrate over configurations such that the the gauge field (and scalars in the supersymmetric case) have a prescribed monopole-like singularity along the given contour. 
It can be seen that 't~Hooft operators in a theory with gauge group $G$  are labeled by representations ${^L}R$ of the \emph{dual} 
group ${^L}G$ \cite{Goddard:1976qe}\cite{Kapustin:2005py}\cite{Kapustin:2006pk}.
According to $S$-duality, the Wilson loop $W_R({\cal C})$ in the theory with gauge group $G$ is then mapped to the 't~Hooft loop $T_{R}({\cal C})$ in the theory with gauge group ${^L}G$, inserted along the same contour ${\cal C}$ and labeled by the representation $R$ of $G$, and vice versa.
 In particular, their quantum expectation values are supposed to be equal upon the replacement (\ref{S-tau-map}). The following table summarizes the action of $S$-duality on Wilson and 't~Hooft operators:
\begin{center}
S-duality \\
\begin{tabular}{|c|c|}
\hline
Gauge group $G$ & Gauge group $^L G$ \\
\hline
$\tau$ &  $^L \tau$ \\
\hline
't~Hooft loop in rep $^L R$ of $^L G$ & Wilson loop in rep $^L R$ of $^L G$  \\
\hline
Wilson loop in rep $R$ of $G$ & 't~Hooft loop in rep $R$ of $G$ \\
\hline
\end{tabular}
\end{center}

Because $S$-duality relates weak and strong coupling dynamics, it is in general hard to perform explicit quantitative tests of the conjecture. However, non-trivial confirmation of the duality may be obtained by studying loop operators which preserve some fractions of the supersymmetries of the theory (in general, combinations of ordinary and superconformal supersymmetries). In this situation, one may in fact be able to obtain \emph{exact} results for their quantum correlation functions, interpolating between weak and strong coupling. The best known example is the 1/2 BPS circular Wilson loop which couples to one of the six scalars field. The expectation value of this operator is exactly captured by a simple Gaussian matrix model, as first conjectured in \cite{Erickson:2000af}\cite{Drukker:2000rr} and proved in \cite{Pestun:2007rz} using localization for the gauge theory on $S^4$. 

Generalizing upon this example, a new large class of supersymmetric Wilson loops has been constructed in \cite{Drukker:2007dw,Drukker:2007qr}. These operators are defined for arbitrary contours on a round $S^3$ in space-time and couple to three of the six scalars. Generically, they are 1/16 BPS. A rather interesting sub-family can be defined by restricting the contours to lie on a great $S^2$ inside $S^3$. The corresponding loop operators are 1/8 BPS and they were conjectured to be exactly captured by the ``zero-instanton sector'' of 2d Yang-Mills theory on $S^2$ \cite{Drukker:2007yx,Drukker:2007qr}. This is in turn related to simple Gaussian matrix models with area dependent couplings \cite{Bassetto:1998sr,Bassetto:1999dg}. The 1/2 BPS circular loop is consistently recovered as a special case, and corresponds to an equator of the $S^2$. Several evidences in favor of the conjecture, both from perturbation theory and from the dual string theory in $AdS_5 \times S^5$, have been presented in \cite{Drukker:2007yx,Drukker:2007qr,Young:2008ed,Bassetto:2008yf,Giombi:2009ms,Bassetto:2009rt}. 

In \cite{Pestun:2009nn}, extending the results of \cite{Pestun:2007rz} to the case of the 1/8 BPS loops on $S^2$, the localization framework for the gauge theory on $S^4$ was used to argue that, for smooth field configurations, the 4d path integral localizes to a 2d field theory which turns out to be closely related to the Yang Mills Hitchin/Higgs theory (YMH) \cite{Moore:1997dj,Gerasimov:2006zt,Gerasimov:2007ap,Nekrasov:2009rc}. For the purpose of computing correlation functions of the 1/8 BPS Wilson loops on $S^2$, this theory was argued in \cite{Pestun:2009nn} to be perturbatively equivalent to pure 2d Yang-Mills theory, and also a natural explanation for the absence of non-trivial 2d instanton contributions (based on the appearance of extra fermion zero modes) was given. The explicit computation of the one-loop determinant for fluctuations normal to the localization locus was left open in \cite{Pestun:2009nn}, but there are reasons to believe that it could be trivial as in the 1/2 BPS case \cite{Pestun:2007rz}, hence the results of \cite{Pestun:2009nn} would essentially support the conjecture of \cite{Drukker:2007yx,Drukker:2007qr}.  

In fact, the localization framework of \cite{Pestun:2009nn} turns out to be rather rich and allows one to establish a more general dictionary between physical observables of the 4d theory which share some supersymmetry with the 1/8 BPS loops and observables of the 2d theory on $S^2$. An example is given by certain local chiral primary operators which can be inserted at arbitrary points on $S^2$: on the 2d theory side, they correspond to insertions of powers of the 2d YM field strength, and exact results for mixed correlation functions of local and Wilson loop operators can be obtained from 2d YM \cite{Giombi:2009ds}. 

Another interesting example, which is the main subject of this paper, is the case of the 1/2 BPS circular 't~Hooft loop operator. By examining the supersymmetry equations of \cite{Pestun:2009nn} which dictate the localization, one can realize that a 1/2 BPS 't~Hooft loop inserted along a great circle of $S^4$ linked to the $S^2$ on which the Wilson loops live is also $Q$-closed, where $Q$ denotes the supercharge used in the localization (one of the four supercharges preserved by the Wilson loops). In other words, the 't~Hooft loop is a particular solution of the supersymmetry equations with a monopole singularity at the center of a solid ball whose boundary is the interesting $S^2$. To rigorously understand how localization works in the presence of the magnetic loop, one should study the full moduli space of solutions of the supersymmetry equations in the presence of the singularity, generalizing the analysis of \cite{Pestun:2009nn} where smooth field configurations were assumed. In this paper we do not perform this analysis, and instead propose a natural conjecture based on the following simple observation: the classical field configuration generated by the 't~Hooft loop, when restricted to the $S^2$, is precisely equivalent to the (unstable) instanton solution of 2d YM labeled by the same quantum numbers of the 't~Hooft loop, i.e. a representation ${^L}R$ of the dual group ${^L}G$. Hence we propose that, in the presence of the 1/2 BPS 't~Hooft loop, the 4d path integral, with possible insertions of $Q$-closed observables, localizes to the path integral of 2d YM around non-trivial unstable instantons. In the case of the minuscule representations\footnote{In a minuscule representation all weights have the same length. For $G=U(N)$, the minuscule representations are the totally antisymmetric representations of arbitrary rank.} there are no complications related to subleading corrections, or so-called ``monopole bubbling" \cite{Kapustin:2006pk,Gomis:2009ir,Cherkis:2007jm}. In this case, we conjecture that the 't~Hooft loop with highest weight ${^L}\lambda$ is captured by the contribution to the 2d Yang-Mills partition function of the unstable instanton labeled by ${^L}\lambda$. For general representations, we expect contributions of instantons associated with shorter weights appearing in that representation. In this paper, we mainly concentrate on the simplest case $G=U(N)$, for which ${^L}G=G$.  Also, here we restrict to the case of imaginary $\tau$ (i.e. $\theta=0$). We leave the study of more general gauge groups and representations, as well as non-zero $\theta$, to future work. 

According to $S$-duality, the expectation value of the 1/2 BPS 't~Hooft loop should be given by the same Gaussian matrix model which captures the 1/2 BPS Wilson loop \cite{Erickson:2000af,Drukker:2000rr,Pestun:2007rz}, with an inverted coupling constant as given by (\ref{S-tau-map}). Recently, this expectation was shown to be consistent with perturbation theory in \cite{Gomis:2009ir}, where a direct one-loop computation of the 't~Hooft loop expectation value was carried out. In this paper, we apply our conjecture to obtain an exact prediction for the vev of the 1/2 BPS 't~Hooft loop in ${\cal N}=4$ SYM, and show that this is indeed precisely given by the Gaussian matrix model with dual coupling constant, as required by $S$-duality. Our conjecture also allows us to derive new exact predictions for correlation functions of the 't~Hooft loop with any number of 1/8 BPS loops on $S^2$, by computing Wilson loop correlators in the 2d theory in the background of an instanton. As an example, we present the result for the correlator of the 't~Hooft loop and the 1/2 BPS Wilson loop at the $S^2$ equator when both operators are labeled by the fundamental representation of $U(N)$, and show that the result is precisely consistent with $S$-duality.

There are several directions in which one may try to complete and extend the present work. 
In this paper we only consider the case when $\tau$ is purely imaginary, i.e.  $\theta = 0$. So one immediate generalization is understand 
the case of general $\tau$, as well as the dyonic Wilson-'t~Hooft operators \cite{Kapustin:2005py,Kapustin:2007wm}.
As mentioned above, to further substantiate our conjecture, one should study the full moduli space of solutions to the localization equations of \cite{Pestun:2009nn} in the presence of the monopole singularity, and rigorously derive the resulting 2d theory. Working out the details and derive new results for Wilson-'t~Hooft correlators for most general gauge groups and choice of representations would also be a natural step in which to test our proposals. Finally, one may include in the story also other physical observables of the 4d theory for which the same localization frameworks applies, for example the local chiral primary operators on $S^2$ studied in \cite{Giombi:2009ds}\footnote{The action of $S$-duality on local operators has been studied in \cite{Intriligator:1998ig,Intriligator:1999ff}, and a perturbative calculation of the local operator-'t~Hooft loop correlator has recently appeared in \cite{Gomis:2009xg}.}. By allowing field configurations which are singular on the interesting $S^2$, one should also be able to include in the same setup supersymmetric surface operators \cite{Gukov:2006jk,Gomis:2007fi,Drukker:2008wr}. This would hopefully give a rich array of new exact results in ${\cal N}=4$ SYM which may be used to further our understanding of the $S$-duality symmetry, and  may be also useful in the context of the holographic duality to string theory in $AdS_5 \times S^5$. Recently, interesting works on loop and surface operators in supersymmetric gauge theories have appeared \cite{Alday:2009fs,Drukker:2009id}, which used the relation between 4d ${\cal N}=2$ gauge theories and Liouville theory uncovered in \cite{Alday:2009aq}. Another recent work \cite{Nekrasov:2009rc} relates 4d susy gauge theory to 2d gauge theory and integrable systems. It would be interesting to find connections between those papers and the present work.  

The paper is organized as follows. In Section \ref{Conv} we set up our notations and conventions. In Section \ref{Abel-S} we briefly review the classical abelian electric-magnetic duality. In Section \ref{BPS-magnetic} we give a general definition of locally BPS 't~Hooft loop operators in ${\cal N}=4$ SYM supported on arbitrary contours, and we evaluate their expectation value in the semiclassical limit.  In particular we discuss regularization and introduce the relevant boundary term which makes the computation finite. In Section \ref{Localize} we review the basic steps of the localization calculation of \cite{Pestun:2009nn}, show that the circular 1/2 BPS 't~Hooft loop solves the relevant supersymmetry equations and state our main conjecture that relates the 't~Hooft loops to the unstable instantons of the 2d theory. In Section \ref{tHooft-2d} and \ref{Wilson-tHooft} we apply our conjecture to derive respectively the 't~Hooft loop expectation value and the Wilson-'t~Hooft correlator from 2d YM.

\subsubsection*{Acknowledgments}
We are grateful to J. Gomis, A. Kapustin and T. Okuda for useful discussions and correspondence. The work of S.G. is supported in part by the Fundamental Laws Initiative Fund at Harvard University and by NSF Award DMS-0244464. The work of V.P. is supported by a Junior Fellowship from the
Harvard Society of Fellows, and grants NSh-3035.2008.2 and RFBR 07-02-00645.

\section{Conventions}
\label{Conv}

We consider  Yang-Mills gauge theory with gauge group $G$ in Euclidean signature. Let $\g$ denote
the Lie algebra of $G$. 
Our convention is that the gauge field $A$ takes value in $\g$, e.g. in the anti-hermitian matrices if $G = \U(N)$. 
By $D_{\mu} = \p_{\mu} + A_{\mu}$ we denote the covariant derivative. The curvature 
is the two-form $F = dA  + A \wedge A$, i.e. $F_{\mu \nu} = [D_{\mu}, D_{\nu}]$. 

In the usual physics notations $D_{\mu} = \p_{\mu} - i A_{\mu}'$ and $F'_{\mu \nu}  = i [D_{\mu}, D_{\nu}]$, 
where $A'$ and $F'$ are represented by Hermitian matrices. So we have the relations
$A = -i A'$ and $F = -i F'$.

The Yang-Mills functional is  
\begin{equation}
  \label{eq:Yang-Mills}
  S_{YM} = -  \frac {1} {\gym^2} \int \tr F \wedge *F  - \frac {i \theta} {8 \pi^2} \int \tr F \wedge F\,,
\end{equation}
which depends on two real coupling constants $\gym$ and $\theta$ which we combine into a complex coupling 
constant 
\begin{equation}
  \label{eq:tau}
  \tau = \frac {\theta} { 2 \pi} + \frac {4 \pi i } {\gym^2}\,.
\end{equation}
In the equation (\ref{eq:Yang-Mills}) the symbol $\tr$ for $\U(N)$ gauge group is the trace in the
fundamental representation. Notice that the bilinear form $\tr (\g, \g)$ is negatively defined, 
so the first term in $S_{YM}$ (\ref{eq:Yang-Mills}) is positive. In coordinate notations 
we have $ \int F \wedge * F = \frac 1 2 \int \sqrt{g} F_{\mu \nu} F^{\mu \nu}$ 
and $\int F \wedge F = \frac 1 2 \int \sqrt{g} \ve_{\mu \nu \rho \lambda} F^{\mu \nu} F^{\rho \lambda}$.

Our Lie algebra conventions are the following. By $T_a$ we denote the generators, or basis elements 
of $\g$, so for $A \in \g$ we write $A = A^{a} T_a$, and we take the coordinates $A^{a}$ to be real. 
We choose metric (Killing form) on $\g$ such 
that the short coroot has length $2$. For example, for $G=\U(N)$ or $G=\SU(N)$ the metric $\la, \ra$ on $\g$ is given by minus
trace in the fundamental representation $\la a,b \ra = -tr_F a b$. 
In the basis $T_a$ the metric has matrix form $g_{ab} = -\tr_F T_a T_b$. We use this metric to raise and lower Lie algebra indices.
The second Casimir operator (Laplacian on group $G$) is defined as 
$C_2 = - T^a T_a = - g^{a b} T_a T_b$, and $C_2$ has positive eigenvalues.  It is eigenvalue in representation $R$ is called $C_2(R)$, 
explicitly $-g^{a b} R(T_a) R(T_b)= C_2(R) 1_{d_R \times d_{R}}$.  For example, for $U(N)$ we have $C_2(F) = N$ and for $\SU(N)$ we have $C_2(F) = \frac {N^2-1}{N}$. 
%Also we define constant 
% $\tr_{R} T_a T_{b} = -C(R) \delta_{a b}$, $R(T_a) R(T_b) = -C_{2} (R) 1$, 
%for example for $\U(N)$ we have $C_{2} (F) = \frac {N}{2}$ while for $\SU(N)$ we have $C_{2}(F) = \frac { N^2 -1 }{ 2 N}$. 
%Of course $C_2(R) d_{R} = C(R) \dim G$.
%Then we normalize the basis elements of $\g$ 
%as $\la T_a T_b \ra = \frac 1 2 \delta_{ab}$, hence in our coordiantes the metric on $\g$ is  $g_{ab} = \frac 1 2 \delta_{ab}$.

\section{Elementary review of abelian S-duality\label{Abel-S}}

\subsubsection*{Electric charge}

Given a contour $C$ and a representation $R$ of $G$ we define
\begin{equation}
  \label{eq:Wilson}
  W_{R} (C) = \tr_{R} \Pexp \oint A_{\mu} dx^{\mu} = \tr_{R} {\mathrm P} e^{-i \oint A'}. 
\end{equation}
Then the partition function in the presence of the Wilson loop $C$  is 
\begin{equation}
  \label{eq:Z-wilson}
   \langle W_{R}(C) \rangle = \frac 1 {Z} \int [\mathcal D A] e^{-S_{YM}} \tr_{R} {\mathrm P} e^{-i \oint A'}.
\end{equation}

Consider the abelian theory $G = U(1)^r$ with coupling constant $\gym$. We take $r = 1$ for brevity and we set $\theta = 0$. It will be elementary 
to generalize to arbitrary $r$ later. 

The $U(1)$ representations are labeled by an integer $R = n \in \BZ$,
and $\tr_n e^{i \alpha} : = e^{n i \alpha}$. To compute (\ref{eq:Z-wilson}) classically 
we need to find the critical point of the exponent in (\ref{eq:Z-wilson})
\begin{equation}
  \label{eq:extremum}
  S_{YM}[A'] + i n \oint A' .
\end{equation}
Of course, we get the usual Maxwell equations with the source
\begin{equation}
\label{eq:max-eq}
  - \frac {2} {\gym^2} D_{\mu} F'_{\mu \nu} + i n J_{\nu}  = 0\,,
\end{equation}
where $J_{\nu}$ is the source supported on the contour $C$. 

%In the Feynman gauge $\p_{\mu} A_{\mu} = 0$
% one gets the Laplace equation $ \Delta A' = i n \frac {\gym^2}{2} J $.
% whose solution is
% \begin{equation}
%   \label{eq:solution-A}
%   A_{\mu} (x)' = -\frac {i \gym^2 n }{8 \pi^2}  \int dy \frac {J_{\mu} (y) } { (x-y)^2}  
% =  -\frac {i \gym^2}{8 \pi^2} \oint_{C}  \frac{ dy_{\mu}} { (x -y)^2} 
% \end{equation}

Let $C$ to be the straight line in the direction $x_0$
located at the origin $x_1 =  x_2 =  x_3 = 0$. Solving (\ref{eq:max-eq}) we get the usual Coulomb law
\begin{equation}
  \label{eq:solutions-A}
  A_0' = - \frac {i \gym^2 n} { 8 \pi r}, \quad E_r' = F_{0 r}' = \frac {i \gym^2 n } { 8 \pi r^2} \frac {x_i}{r}.
\end{equation}
Here $r$ is the 3d distance from the origin and 
$E_{r} = F_{0r} = F_{0i} \frac {x^{i}} {r}$ is the radial component of the electric field strength. 

The energy of the point electric charge diverges. To regularize it, we 
introduce a UV cut-off small distance $\ve$ and delete a solid tube of radius $\ve$ 
surrounding the contour $C$. 

The contribution of the action term in (\ref{eq:extremum}), evaluated per unit time, gives 
\begin{equation}
  \label{eq:direct-integral}
  S_{YM}[A_{cl}^{el},n] = \frac 1 {\gym^2} \int_{\ve}^{\infty} 4 \pi r^2 dr (F'_{0r})^2  = - \frac { n^2 \gym^2}{16 \pi \ve }\,.
\end{equation}
The contribution of the source term $i n \oint A'$  is
\begin{equation}
  \label{eq:source}
  i n  A'_{cl}|_{r = \ve} =  \frac {\gym^2 n^2}{ 8 \pi \ve }\,.
\end{equation}
As a result we get that classical regularized vev of the Wilson loop per unit length
is computed as $e^{-\mathcal{E}_{elec}}$ where
\begin{equation}
  \label{eq:total}
 \mathcal{E}_{elec} = S_{YM}[A_{cl}]+ in A'_{cl} = \frac{ \gym^2 n^2} { 16 \pi \ve}\,.
\end{equation}

\subsubsection*{Magnetic charge} 
Now we consider 't~Hooft loop operator in the $\U(1)$ theory with coupling constant $\gym.$ 
The classical Maxwell equations in the absence of source terms are invariant under 
exchange of electric and magnetic fields. Hence we find the field strength associated to the magnetic 
charge is given by
\begin{equation}
  \label{eq:magnetic}
\begin{aligned}
 F'_{0j} = 0 \\
 F'_{jk} =  \frac{m} {2}  \ve_{ijk}  \frac {x_i}{r^3},
\end{aligned}
\end{equation}
where $m$ is yet an arbitrary constant and the factor $1/2$ is introduced for convenience. 
The constant $m$ is quantized, of course, as can be easily seen by integrating the two form $F$ over 
a spherical surface $S^2$ surrounding the magnetic charge. One can see that $m$ has the meaning 
of the first Chern class for the gauge bundle restricted to $S^2$
\begin{equation}
  \label{eq:chern-class}
 m = \frac {i} {2 \pi} \int_{S^2} F,
\end{equation}
and hence $m$ is an arbitrary integer. This integer is the magnetic charge of the 't~Hooft loop operator. 

The action for the magnetic charge diverges like in the case of the electric charge. To compute 
the regularized action we integrate outside the tubular neighborhood of radius $w$ surrounding
the magnetic charge. Then we get 
\begin{equation}
  \label{eq:direct-integral-magnetic}
  \mathcal{E}_{mag} = S_{YM}[A^{mag}_{cl},m] = \frac 1 {\gym^2} \int_{w}^{\infty} 4 \pi r^2 dr (F')^2  =  \frac { \pi m^2 }{\gym^2 }\,.
\end{equation}

\subsubsection*{Abelian S-duality}

Under S-duality the Wilson loop is mapped to the 't~Hooft loop while the coupling constants 
are related as
\begin{equation}
  \label{eq:coupling}
   \gym^2 \mapsto \frac {16 \pi^2} { \gym^2}.
\end{equation}
Clearly,   the energies (\ref{eq:direct-integral}) and (\ref{eq:direct-integral-magnetic}) 
coincide under (\ref{eq:coupling}) and $n \mapsto m$.

%We can describe this mapping as 
%\begin{equation}
%  \label{eq:abelian-S-duality}
%  F \mapsto *F. 
%\end{equation}

\section{Locally BPS 't~Hooft operator\label{BPS-magnetic}}

In the $\CalN=4$ Yang-Mills it is customary to study Wilson loop operators coupled to scalar fields \cite{Maldacena:1998im}
\begin{equation}
  \label{eq:local-BPS}
  W_{R}(C) = \frac{1}{d_{R}} \tr_{R} \Pexp \oint A_{\mu} dx^{\mu} + i \theta^{A}(s) \Phi_{A} ds.
\end{equation}
Here $\theta^{A}(s)$ specifies couplings to the scalar fields $\Phi_A$, $A = 1\dots 6$, of $\CalN=4$ SYM theory. 
If $\theta^A \theta^A = 1$ then the operator (\ref{eq:local-BPS}) is called locally BPS, because 
for any point $x$ on the contour $C$ one can find 8 supercharges  $Q_{\alpha}(x)$ which locally 
annihilate Wilson loop~(\ref{eq:local-BPS}) at the point $x$. For special choices of contour and of $\theta^A(s)$ one can obtain operators which globally preserve some supercharges \cite{Zarembo:2002an,Drukker:2007qr}. The well known 1/2 BPS case is obtained by $\theta^A(s)=const$, and by taking the contour to be a circle (or a straight line). 

It is also elementary to check that in the leading order of perturbation theory the Wilson loop (\ref{eq:local-BPS}) on smooth contour $C$ is finite 
if and only if $\theta^A(s) \theta^A(s) = 1$. 
The propagators for the gauge and scalar fields on $\BR^4$ are (we choose Feynman gauge $\p_{\mu} A_{\mu} = 0$)
\begin{equation}
  \label{eq:propagators}
  \begin{aligned}
   &G_{\mu \nu}^{a b} := \la A_{\mu}^a A_{\nu}^b \ra =\frac{ \gym^2}{8 \pi^2} \frac {g^{ab} g_{\mu \nu}}{(x-y)^2} \\
   &G_{A B}^{a b} :=  \la \Phi^a_A \Phi_{B}^b \ra =  \frac{ \gym^2}{8 \pi^2} \frac {g^{ab} \delta_{AB} }{(x-y)^2}.
  \end{aligned}
\end{equation}
Then in the leading order we get
\begin{equation}
  \label{eq:leading-Wilson-loop}
  \la W_{R}(C) \ra  =  1 - \frac{\gym^2} {16 \pi^2} C_{2}(R) G^{(2)}(C) \,,
\end{equation}
where we have denoted
\begin{equation}
  \label{eq:GC-notation}
 G^{(2)}(C) =  \oint ds \oint  ds' (  \frac { \dot x^{\mu}(s) \dot x_{\mu}(s')  
 -\theta^A(s) \theta_A (s') } { (x(s)-x(s'))^2 }).
\end{equation}
 We  notice that the contour shape dependent functional $G^{2}(C)$ 
is negative and conformally invariant. In the case of circle and constant $\theta^A$ we get $G^{(2)}(C) = -2 \pi^2 $. 
In the abelian case, for $G = \U(1)^{r}$, it is elementary to compute exact expectation value of the Wilson loop (\ref{eq:local-BPS}) because
the path integral is Gaussian. The Wilson loop (\ref{eq:local-BPS}) reduces 
to 
\begin{equation}
  \label{eq:wilson-abelian}
  W_{R}(C)^{abel} = \frac{1} {d_R} \sum_{\alpha \in \mathrm{irreps}(R) } e^{ i  w^{\alpha}_a \oint A^a_{\mu} dx^{\mu} + i \theta^{A}(s) \Phi^a_{A} ds}.
\end{equation}
Here index $\alpha$ runs over irreducible representations in the decomposition $R = \sum_{\alpha} R_{\alpha}$. 
Each irreducible representation of abelian group $G$ is one-dimensional and is defined by its weight $w$, which is a one-form on $\g$, 
by the rule that the an element of $G$ of the form $e^{A^a T_a}$ is represented by a complex number $e^{i A^a w_{a}}$. 
Computing the Gaussian integral with insertion (\ref{eq:wilson-abelian}) we get 
\begin{equation}
\begin{aligned}
  \label{eq:wilson-abelian-expect}
  \la W_{R}(C)^{abel} \ra  &= \frac{1}{d_R} \sum_{\alpha \in \mathrm{weights}(R)} 
e^{ - \frac 1 2 \la   w_a \oint (A^a_{\mu} dx^{\mu} + i \theta^{A}(s) \Phi^a_{A} ds)   (w_b \oint A^b_{\nu} dx^{\nu} + i \theta^{B}(s) \Phi^b_{B} ds)  \ra } 
=
\\ 
&= \frac {1}{d_R} \sum_{\alpha \in \mathrm{weights}(R)} e^{-  {\frac {\gym^2} {16 \pi^2} \la w^{\alpha}, w^{\alpha} \ra   G^{(2)}(C)}}.
\end{aligned}
\end{equation}
Irreducible representations of $G=\U(1)^r$ are labeled by $r$-dimensional integer vector $\vec{n} \in \BZ^{r}$. 
If metric on $\g$ is fixed as minus trace in the fundamental representation then $ \la w, w \ra = \vec{n}^2$
for weight $w$ associated to representation $\vec{n}$.

The Wilson loop operator (associated to an electric charge)  is the usual operator defined as a functional on the space of fields. To compute
expectation value (or correlation functions) for Wilson loop operator, one just insert the corresponding 
functional under the sign of the path integral. On the other hand, the 't~Hooft operator (associated to 
magnetic charge, or monopole) is a disorder operator, 
defined by a prescribed singularity for the fields \cite{Kapustin:2005py,Kapustin:2006pk}. 
To compute expectation value or correlation functions for 't~Hooft operator, one actually changes the 
definition of the path integral itself. Instead of integrating over arbitrary smooth fields on space-time,
we require the fields to be smooth everywhere except at the location of the disorder operator, where the fields 
are required to have the prescribed singular behavior. 

More concretely, the 't~Hooft operator is defined as follows \cite{Kapustin:2005py,Kapustin:2006pk,Witten:2009mh}. 
For the gauge group $G$ we choose group homomorphism $\rho: \U(1) \rightarrow G$. Such homomorphisms $\rho$
are labeled by the coweights of $G$, or, equivalently, by the weights of the dual group $^L G$. Given $\rho$ 
and the contour $C$ the 't~Hooft operator is defined by asking the gauge fields to have singularity 
near $C$ like the image under $\rho$ of the basic $\U(1)$ monopole (\ref{eq:magnetic}).

We are particularly interested in the partially BPS supersymmetric 't~Hooft loops in the $\CalN=4$ super Yang-Mills. 
Similarly to the supersymmetric Wilson loop operator, which couples to the scalar fields in the $\CalN=4$ super Yang-Mills, we also turn on coupling to the scalar fields for the supersymmetric 't~Hooft operator.

Now we define locally BPS 't~Hooft loop operator by generalizing the definition of 1/2 BPS 't~Hooft loop operator supported 
on a straight line \cite{Kapustin:2005py} (here we consider the $\Re\tau=0$ case). Given a smooth, not self-intersecting contour $C$ and smooth 
couplings $\theta^{A}(s)$, $s \in C$ such that $\theta(s)^2 = 1$,  we require
that the gauge field and the scalar field have the following singularity in the neighborhood of $C$
\begin{equation}
  \label{eq:def-local-BPS-t-Hooft}
\begin{aligned}
  F_{kl}(y) &= \frac{1}{2} \ve_{ijkl }  \frac{d x^i} {ds} \frac{ (y_j-x_j)} {|y-x|^3} T_{\vec{m}} + O(1) \\
  \Phi_{A }(y) &= \frac {\theta^A(s)} {2 |y-x|} T_{\vec{m}} + O(1), \quad \text{in the limit} \quad  |y-x| \to 0  \\
   T_{\vec{m}} &:=  -i \diag(m_1, \ldots, m_N)\,,
\end{aligned}
\end{equation}
where for each point $y$ in the neighborhood of $C$, the point $x \in C$ is the point closest to $y$. 
If we consider normal hyperplane $\BR^{3}_x$ for each point $x \in C$, 
the fields $F$ and $\Phi = \Phi_A \theta^A$ approximately satisfy Bogomolny equation \cite{Bogomolny:1975de,Witten:2009mh} in the infinitesimal neighborhood of $x$ 
\begin{equation}
  \label{eq:Bogomolny}
  *_{\BR^3} F  + d \Phi = 0,
\end{equation}
hence the singularity (\ref{eq:def-local-BPS-t-Hooft}) defines a locally BPS 't~Hooft operator. The globally supersymmetric 1/2 BPS 't~Hooft loop is given by (\ref{eq:def-local-BPS-t-Hooft}) with $\theta^A=const$ and $C$ straight line or circle. 

The expectation value 
and correlation functions of 't~Hooft operator are defined by taking the path integral over all fields 
with the asymptotics (\ref{eq:def-local-BPS-t-Hooft}). In the semiclassical limit, the main contribution 
to the path integral is given by the critical point of the action, i.e. by a classical configuration 
which satisfies the equations of motion and has the required asymptotics (\ref{eq:def-local-BPS-t-Hooft}). 
Clearly, the action evaluated on such configuration will diverge in the region close to the contour $C$. 
The difference with the corresponding computation for locally BPS Wilson loop case is that in the Wilson case the 
divergent contributions coming from the action for gauge field was of the same magnitude but of opposite sign 
as the contribution coming from the  scalar field. In the 't~Hooft case both contributions (for the gauge field and for the scalar field) are of the 
same sign and are not cancelled. 
This puzzle, which naively seems to violate the  S-duality (say in the abelian case, where classical computation is supposed to be exact), 
is easily resolved by recalling that when we do semiclassical computation in the Wilson case, 
to the Yang-Mills action evaluated on classical solution we need to add  the source term (\ref{eq:extremum}).
Similarly, in the locally BPS Wilson case, we need to add the source term for $\Phi$ when we compute classical expectation value. 
In the 't~Hooft case there is no natural source for the magnetic field, and it is not actually needed in order for 
abelian $S$-duality to work. Indeed, for gauge fields one can see
that $S_{YM}(^L \gym, ^L G ;F_W^{cl}) + S_{source}(A_W^{cl}) = S_{YM}(\gym, G; F_T^{cl})$, where $A_W^{cl}, F_{W}^{cl}$ and $F_{T}^{cl}$ are fields
created by Wilson loop or 't~Hooft loop respectively.  The naive divergence problem of locally BPS 't~Hooft loop 
and the naive disagreement with the dual locally BPS Wilson loop comes actually from the scalar sector, for the simple
reason that in the Wilson case we have taken into account contribution of the source term for $\Phi$, but in the 't~Hooft
case we have not. Moreover since $S_{source}(\Phi^{cl}) = - 2 S_{YM}(\Phi^{cl})$, just like for the gauge field, we have
that $S_{source}(\Phi^{cl}) + S_{YM}(\Phi^{cl}) = - S_{YM}(\Phi^{cl})$. So our conclusion is that 
the natural way to resolve this puzzle about divergence and mismatch with $S$-duality  is just to add a source term for the field $\Phi$, 
chosen such that it creates configuration (\ref{eq:def-local-BPS-t-Hooft}), to the definition of 't~Hooft loop operator.

For computational purposes we try to give the following more detailed definition of locally BPS 't~Hooft loop. We try to 
give a general definition for contour of arbitrary shape and space-time manifold $M$ equipped with arbitrary Riemannian metric. 
For a smooth not self-intersecting
contour $C$ let $D(C, \ve)$ denote a solid tubular  neighborhood of the contour
$C$ of size $\ve$
\begin{equation}
  \label{eq:D-C}
D(C, \ve) = \{ x \in M | \mathrm{distance}(x, C) < \ve \}.
  \end{equation}
Then  $M(C,\ve) = M \setminus D(C,\ve)$ is a four-dimensional manifold with a boundary. We  call this boundary $\Sigma_3(C,\ve) = \p D(C,\ve) = 
- \p M(C,\ve)$. 
In the path integral we integrate over all field configurations in the 4d bulk space $M(C, \ve)$. For the gauge field we fix 
Dirichlet boundary conditions on $\Sigma_3(C,\ve)$ as given by classical configuration which satisfies (\ref{eq:def-local-BPS-t-Hooft}). 

For the scalar fields we fix Neumann boundary conditions on $\Sigma_3(C, \ve)$ as defined by (\ref{eq:def-local-BPS-t-Hooft}), 
or, equivalently, we can insert source term for the field $\Phi$ with support on the boundary $\Sigma_3(C, \ve)$.
While specifying boundary conditions for the fields in the form (\ref{eq:def-local-BPS-t-Hooft}) we 
break the gauge group $U(N)$ to $U(1)^{r}$ on the boundary. In other words, when we factorize the path 
integral over gauge transformations we require a gauge transformation $g(x)$ to be a smooth $G$-valued 
function on $M(C,\ve)$ with boundary conditions on $\Sigma_{3}(C,\ve)$ specified by restricting $g(x)$ to the maximal 
torus $T \in G$ for $x \in \Sigma_{3}(C,\ve)$. 
For closed contour, the 3d manifold $\Sigma_3(C,\ve)$ has topology $S^1 \times S^2$, and for sufficiently small $\ve$ 
it can be naturally given the structure of foliation. Namely, for each point $s \in C$ define the
two-manifold $\Sigma_{2}(s,C,\ve) \subset \Sigma_{3}(C, \ve)$ as a set of point in $\Sigma_{3}(C, \ve)$ which 
are located at the distance $\ve$ from $s$ (and $\ve$ is minimal possible distance). Then $\Sigma_{3}(C,\ve)$ 
is represented as a $S^2$-fiber bundle over $C$, where for each point $s \in C$ the $S^2$-fiber is $\Sigma_{2}(s,C,\ve)$. In the following, we will employ the short-hand notation $M(C,\ve)=M_{\ve}$, $\Sigma_{3}(C, \ve)=\Sigma_{3}$, $\Sigma_{2}(s,C,\ve)=\Sigma_{2}(s)$.
Given the structure of the fiber bundle $\Sigma_{2}(s) \to \Sigma_{3} \to C$, there is a natural coordinate $s$ on 
$\Sigma_{3}$ induced by length parameter $s$ on $C$, and the associated one-form $ds$. 
Also, the scalar couplings $\theta^{A}(s)$ could be pulled back on $\Sigma_{3}$ from $C$.

Given the above geometrical definitions, one natural way to write down the  source term for the field $\Phi$ 
is the boundary action on $\Sigma_3$ of the form\footnote{We thank A. Kapustin for a useful discussion.}
\begin{equation}
   \label{eq:boundary-action}
 \frac{2}{\gym^2} \tr  \int_{\Sigma_{3}} F \wedge \Phi_A \theta^A \wedge ds.
\end{equation}
This boundary action can be interpreted as a source term for the field $\Phi$ after we integrate over gauge fields, 
so that $F$ becomes proportional to the volume form on the $S^2$ fibers. 
Such boundary term is natural from the point of view of Bogomolny equations. The YM action coupled to the scalar field 
$\Phi$ on a 3d manifold $M_3$ with boundary $\partial M_3$ is the square of the equation (\ref{eq:Bogomolny}) up to the boundary term 
\begin{equation}
  \label{eq:YM-actionR3}
  -\tr \int_{M_3} ( *F + D \Phi) \wedge *(*F + D \Phi) = -\tr \int_{M_3} (F \wedge * F + D\Phi \wedge * D \Phi ) + 2 \tr \int_{\p M_3} F \wedge \Phi\,.
\end{equation}
The total bulk and boundary action is then
\begin{multline}
  \label{eq:S-bulk-boundary}
   S_{YM} + S_{boundary} = - \frac {1}{ \gym^2} \tr  \int_{M(\ve)} F \wedge * F + d \Phi_A \wedge * d \Phi_A  
+ \frac{ 2}{\gym^2} \int_{\Sigma_{3}} \tr F \wedge \Phi_A \theta^A \wedge ds,
\end{multline}
and by (\ref{eq:YM-actionR3}) it clearly vanishes per unit length for the 1/2 BPS 't~Hooft line.

Now to compute expectation value of locally BPS 't~Hooft loop semiclassically it is enough to evaluate the total 
action (\ref{eq:S-bulk-boundary}) on a classical configuration with asymptotics (\ref{eq:def-local-BPS-t-Hooft}). 
These classical fields can be easily found. We make the computation on $\BR^{4}$ to make it more transparent. Then we have
\begin{equation}
  \label{eq:classical-fields-tHooft}
\begin{aligned}
 &\Phi^{cl}_{A}(y) = \frac {T_{\vec{m}}} {2 \pi} \oint_{C} 
\frac {\theta^{A}(s)\, ds} { (y - x(s))^2} \\
F^{cl}_{kl} = \frac 1 2 \ve_{ijkl} (\p_i b_j^{cl} - \p_j b_i^{cl}), \quad \text{where} \quad 
&b^{cl}_i(y) = \frac {T_{\vec{m}}} {2 \pi} \oint_{C} \frac { dx_i }{ (y-x(s))^2}\,.
\end{aligned}
\end{equation}
Since the configurations (\ref{eq:classical-fields-tHooft}) solves the equations of motion  $\Delta \Phi = 0, d F  = 0, d * F = 0$ 
in the bulk $M_{\ve}$, we can evaluate (\ref{eq:S-bulk-boundary}) integrating by parts and reducing the integral to the boundary $\Sigma_{3}$.
Classically for abelian configurations we have
\begin{equation}
  \label{eq:on-shell}
\begin{aligned}
  \int_{M_{\ve}} F \wedge * F  = \int_{M_{\ve}} db \wedge * db = \int_{M_{\ve}} d (b \wedge * db) = - \oint_{\Sigma_3} b \wedge * db \\
  \int_{M_{\ve}} d \Phi \wedge * d \Phi = -\oint_{\Sigma_3} \Phi \wedge * d\Phi,
\end{aligned}
\end{equation}
so
\begin{equation}
  \label{eq:classical-boundary}
  S_{YM} + S_{boundary} = \frac {1} {\gym^2} ( \tr \oint_{\Sigma_3} b \wedge *_4 db + \Phi \wedge *_4 d\Phi - 2 \Phi ds \wedge *_4 db).
\end{equation}
Now we can plug in the classical solution (\ref{eq:classical-fields-tHooft}) into (\ref{eq:classical-boundary})
and take the limit $\ve \to 0$. In this limit we can average the values of fields $b$ and $\Phi$ over each fiber $\Sigma_2(s)$ using that
\begin{equation}
  \label{eq:ep-to-0}
  \begin{aligned}
  &\int_{\Sigma_2(s)} *_4 db = 2 \pi T_{\vec{m}}  + O(\ve) \\
  &\int_{\Sigma_2(s)} *_4 d\Phi_A = 2 \pi T_{\vec{m}} \theta_A (s)ds  + O(\ve)\, ds,
  \end{aligned}
\end{equation}
so we get 
\begin{multline}
  \label{eq:S-YM-S-bound-cont}
  S_{YM} + S_{boundary} = - \frac { 2\pi}{\gym^2} \lb 
\oint_{C} \tr T_{\vec{m}} b_{cl}(s)  - \oint_{C} \tr T_{\vec{m}}  \Phi^{A}_{cl}(s) \theta_A(s) ds \rb = 
\frac{1}{\gym^2} \vec{m}^2 G^{(2)}(C),
\end{multline}
and finally, 
the classical expectation value of locally BPS 't~Hooft on arbitrary contour $C$ is given by 
\begin{equation}
  \label{eq:result}
  \la T_{R}(C) \ra = \exp ( - (S^{cl}_{YM} + S^{cl}_{boundary})) = \exp ( - \frac{1}{\gym^2} \vec{m}^2 G^{(2)}(C))\,.
\end{equation}

We observe that the result for locally BPS 't~Hooft loop 
clearly agrees with the $S$-dual contribution to the Wilson loop (\ref{eq:wilson-abelian-expect})
under replacement $\gym^2 \to \frac {16 \pi^2}{ \gym^2}$.

\section{Localization of 4d $\CalN=4$ SYM to the 2d theory \label{Localize}}
We consider the same geometrical setup of \cite{Pestun:2009nn} where, extending the work \cite{Pestun:2007rz}, it was shown 
how to use localization in the context of the 1/8 BPS Wilson loops of \cite{Drukker:2007dw,Drukker:2007yx,Drukker:2007qr} to obtain from $\CalN=4$ SYM the two-dimensional theory on $S^2$, which was called in \cite{Pestun:2009nn}  \emph{almost} 2d Yang-Mills theory. This theory is related to the Yang-Mills-Higgs theory \cite{Gerasimov:2006zt,Gerasimov:2007ap,Moore:1997dj,Nekrasov:2009rc}. 

In  \cite{Pestun:2009nn} a set of supersymmetric equations was derived from the appropriate fermionic symmetry of the Wilson loop operators, and there it was shown that the smooth solutions of these equations are parameterized
by two-dimensional data, i.e. by certain  field configurations on $S^2$.
 It was also mentioned in \cite{Giombi:2009ds} that within the same setup one can consider 
the solutions of those supersymmetric equations with singularities which correspond to the insertion 
of 't~Hooft loop operators. 
 
Now we briefly review the construction in \cite{Pestun:2009nn}. We take the space-time 
to be the four-sphere $S^4$, which can be interpreted as the one-point compactification
of $\BR^{4}$. Then we represent the $S^4$ as a warped $S^2 \times S^1$ fibration over 
an interval $I$, such that the metric takes the form 
\begin{equation}
  \label{eq:S4-metric}
  ds^2 =  d\xi^2  + \sin^2 \xi ( d\theta^2 + \sin^2 \theta d \phi^2)   + \cos^2 \xi d\tau^2. 
\end{equation}
The $\xi \in [0, \pi/2]$ is the coordinate on the interval $I$, the $\tau$ is the coordinate 
on $S^1$ fiber,  and $(\theta, \phi)$ are the usual polar coordinates on the $S^2$ fiber. 
At $\xi = 0$ the $S^2$ fiber shrinks to zero size, at $\xi = \pi/2$ the $S^1$ fiber shrinks 
to zero size. The relevant 1/8-BPS Wilson loops studied in \cite{Drukker:2007yx,Drukker:2007qr,Drukker:2007dw,Giombi:2009ds,Giombi:2009ms,Bassetto:2009rt,Bassetto:2008yf} are located at the largest $S^2$ fiber at $\xi = \pi /2$.

The fermionic charge $Q$ used in the localization computation \cite{Pestun:2009nn} squares
to a combination of a  $\U(1)$ rotation along the $S^1$ direction $\tau$ and a rotation in a $\U(1)$ subgroup 
of the $\SO(6)$ R-symmetry of $\CalN=4$ SYM. By the usual arguments the field theory localizes 
to the equations $Q \Psi = 0$ where $\Psi$ are fermionic fields of the theory. In the $\CalN=4$ theory 
one gets sixteen equations, one for each component of $\Psi$. Then it can be seen that nine equations 
tell us that all fields are covariantly constant along the $S^1$ fiber. At this step the 
4d theory localizes to 3d theory on the $S^1$ quotient of the $S^4$. This quotient has topology and the natural 
metric of the solid three-dimensional ball  with boundary, which we denote as $D^3$. The metric on $D^3$ is given by
the first two terms in  (\ref{eq:S4-metric}), which is just the metric on a three-dimensional semi-sphere.

To write down the susy equations, it is actually convenient to make a smooth Weyl transformation of the metric 
on $D^3$ such that the resulting metric is the standard flat metric on the solid ball. Explicitly, after the rescaling, 
that 4d metric on the warped fibration  $D^3 \times_{\tilde w} S^1$ becomes 
\begin{equation}
  \label{eq:metric-after-rescaling}
  ds^2 = dx_i dx_i + \frac 1 4 (1 - x^2)^2d\tau^2, \quad i = 2,3,4.
\end{equation}
Here the $x^i$, $i=2,3,4$ are the standard flat coordinates on $D^3=\{ x_i \in \BR |  x_i x_i  \leq 1 \}$. Because of the conformal 
symmetry of the equations we are free to do such rescaling.

The remaining seven supersymmetric equations are 3d equations on $D^3$ for
the 3d gauge field and five scalar fields (one of six scalar fields of $\CalN=4$ SYM does not appear
in the 3d equations). The equations are invariant under a diagonal $SO(3)$ subgroup of $SO(3)_{Lorentz}\times SO(3)_R$, where $SO(3)_R$ is a subgroup of the $SO(6)_R$ $R$-symmetry group. The scalars which transform under $SO(3)_R$ are denoted by $\Phi_{6}, \Phi_{7}, \Phi_{8}$ (these are the three scalars which couple to the 1/8 BPS Wilson loops \cite{Drukker:2007dw,Drukker:2007yx,Drukker:2007qr}). 
The remaining two scalar fields are labeled as $\Phi_{5}$
and $\Phi_{9}$.\footnote{The fields $\Phi_{6}, \ldots, \Phi_{9}$ are exactly the original scalar fields of the $\CalN=4$ theory, 
but the field $\Phi_5$ here denotes a twisted combination $\Phi_5 = \sin \tau \Phi_0'  + \cos \tau \Phi_5'$, where
$\Phi_0'$ and $\Phi_9'$ stand for the original scalars of $\CalN=4$ SYM. The orthogonal twisted field $\Phi_0 = -\sin \tau \Phi_5' + \cos \tau \Phi_0$ 
does not couple to the 3d equations. In the absence of a $\theta$-angle, and if singular configurations on the $S^2$ are not allowed, the remaining nine supersymmetry equations are solved by $\Phi_0=0$ and $F_{\tau i}=0$.}

Now we quote the relevant 3d equations from  \cite{Pestun:2009nn}
\begin{equation}
  \label{eq:final-top-equations-original-variables}
\begin{aligned}
& \begin{split}  - (1-x^2) D_k \Phi_9  - \frac 1 2 F_{ij} \ep_{ijk} (1+x^2) + \frac 1 2
[ \Phi_{i+4}, \Phi_{j+4}] \ep_{ijp}(\delta_{pk} - x^2 \delta_{pk} + 2 x_p x_k)  \\
 - [\Phi_5, \Phi_{j+4}](\delta_{jk} + x^2 \delta_{jk} - 2x_j x_k) + 2 \Phi_9 x_k = 0 \,,
\end{split}\\
&\begin{split} 
 [\Phi_9, \Phi_{i+4}] (\delta_{ik} + x_i x_k - x^2
 \delta_{ik})  - D_i \Phi_5(\delta_{ik} - x_i x_k + x^2 \delta_{ik}) +
 2 \Phi_5 (1-x^2)^{-1} x_k \\ + D_i \Phi_{j+4} ( \ep_{ijk} - x_i
 x_p \ep_{jpk} - x_j x_p \ep_{ipk}) - 2 \Phi_{i+4} \ep_{ijk} x_j e_{k+4} = 0 \,,
\end{split} \\
& \begin{split}
  [\Phi_9, \Phi_5](1-x^2) + D_i \Phi_{j+4}(\delta_{ij} +
 \delta_{ij} x^2 - 2 x_i x_j) - 2 \Phi_{j+4} x_j = 0 \,.
\end{split}
\end{aligned}
\end{equation}
 
It is convenient to represent the $\Phi_{i+4}$, $i = 2,3,4$ scalar fields as three components of adjoint valued one-form.  
Then the gauge field and the adjoint valued one-form can be combined into a complexified connection, 
while the remaining two scalars can be combined into a complexified scalar. At the origin of $D^3$ the equations 
take the form of the extended Bogomolny equations \cite{Kapustin:2006pk,Witten:2009mh} which generalize the usual Bogomolny 
equations by doubling the number of fields 
%The limit of the equations (\ref{eq:final-top-equations-original-variables}) near the origin is given by the extended Bogomolny equations,
%hence we can describe the singularity using these simpler equations
\begin{equation}
\begin{aligned}
\label{eq:ext-Bogom}
 -*(F - \Phi \wedge \Phi) - d_A \Phi_9 + [\Phi, \Phi_5] &= 0\,,  \\
* d_A \Phi - d_A \Phi_5 - [\Phi, \Phi_9] &= 0\,, \\
d_A * \Phi + [\Phi_9, \Phi_5] &= 0.
\end{aligned}
\end{equation}
Here $\Phi$ denotes the adjoint valued one-form whose components are $\Phi_{6}, \Phi_{7}, \Phi_{8}$, and $*$ is the three-dimensional Hodge star. 

The equations (\ref{eq:final-top-equations-original-variables}) look complicated, however their detailed analysis 
in the absence of singularities is possible, and one gets the moduli space which is parameterized by 
certain data on the two-dimensional boundary \cite{Pestun:2009nn}, roughly speaking by the fields of almost 2d Yang-Mills theory.

Now we want to insert a supersymmetric 't~Hooft loop running along the $S^1$ fiber at the origin ($0 \leq \tau < 2 \pi$,  $x_2 = x_3 = x_4 = 0$).
This is equivalent to introducing a singularity at the origin of the prescribed form 
into the solutions of the equations (\ref{eq:final-top-equations-original-variables}). Let us first look at the simplified equations (\ref{eq:ext-Bogom}) close to the origin of $D^3$. To introduce the conventional BPS monopole singularity, we can actually set to zero the fields ($\Phi, \Phi_5$) in (\ref{eq:ext-Bogom}).
Then one is left with the classical Bogomolny equation for BPS monopole 
\begin{equation}
  \label{eq:Bog}
  *F + d_A \Phi_9 = 0.
\end{equation}
Effectively abelian solutions with singularity at the origin for the $\U(N)$ gauge group are easily described 
by coupling the $\U(1)$ monopole (\ref{eq:magnetic}) with the scalar field $\Phi_9$ and picking up a homomorphism $\rho: \U(1) \to G$. 
Explicitly, for $\rho$ represented by $N$-tuple $(m_1,\ldots,m_N)$ we ask the fields to have singularity near the origin of the form 
\begin{equation}
\begin{aligned}
  \label{eq:tHooft-loop-singularity}
  F_{jk} = \frac{1}{2} \ve_{ijk} \frac {x_i} {r^3} T_{\vec{m}}\,, \\ 
  \Phi_9 = \frac {1} {2 r} T_{\vec{m}},
\end{aligned}
\end{equation}
where $ r = \sqrt{x_i x_i}$.

%///Somewhere need to say about ``regularization of 't~Hooft loop'' using the boundary term on the tube surrounding the loop of the form
% $\int_{S^2} {F \wedge \Phi}$ to get finite expectation values 

After understanding the solution (\ref{eq:tHooft-loop-singularity}) for the simplified equations, it is elementary 
to write down the effectively abelian (breaking $U(N)$ to $U(1)^N$) solution of the full system (\ref{eq:final-top-equations-original-variables})
with the same kind of singularity. Namely, just set again to zero the fields  $\Phi_{5}, \ldots, \Phi_{8}$. Then equations (\ref{eq:final-top-equations-original-variables})
are consistently reduced to 
\begin{equation}
\label{eq:full-reduced}
 (1+x^2)  \frac 1 2 F_{ij} \ep_{ijk}   + D_k (\Phi_9 (1-x^2)) = 0\,.
\end{equation}
Now one can check that the solution (\ref{eq:tHooft-loop-singularity}) satisfies the equations (\ref{eq:full-reduced}), 
and hence it also solves the full system (\ref{eq:final-top-equations-original-variables}) (provided we keep $\Phi_{5}, \ldots, \Phi_{8} = 0$). Of course, (\ref{eq:tHooft-loop-singularity}) is nothing but the singularity associated with a 1/2 BPS circular 't~Hooft loop. 

So far we have presented just one point on the moduli space of solutions of susy equations (\ref{eq:final-top-equations-original-variables})
with a prescribed singularity at the origin. Let us call this moduli space $\CalM_{\vec{m}}$, and the point corresponding to the solution (\ref{eq:tHooft-loop-singularity}) as a reference point $p_{\vec{m}} \in \CalM_{\vec{m}}$. 

To complete the localization analysis, one would like to find the complete moduli space $\CalM_{\vec{m}}$ and map it to the
two-dimensional data on the boundary $S^2$, similarly to what has been done in \cite{Pestun:2009nn} for smooth solutions. 
The four-dimensional path integral is then reduced to the two-dimensional path integral over $\CalM_{\vec{m}}$, or, 
equivalently, over the  the boundary data on $S^2$.

We leave the detailed analysis of the equations and of $\CalM_{\vec{m}}$ for future study. 
In the present work we just look at point $p_{\vec{m}} \in \CalM_{\vec{m}}$ from the following perspective. 
Namely, we observe that the gauge field (\ref{eq:tHooft-loop-singularity}) restricted to the boundary sphere $S^2$ has
precisely the form of the (unstable) instanton in the two-dimensional Yang-Mill labeled by N-tuple $m_1,\ldots, m_N$. Recalling that in the absence of any singularities 
the $\CalN=4$ SYM four-dimensional path integral has been argued to reduce to the zero-instanton sector of two-dimensional Yang-Mills \cite{Pestun:2009nn},
it is tempting to conjecture that \emph{integration over $\CalM_{\vec{n}}$ is equivalent to the corresponding (unstable) instanton contribution 
to the partition function of two-dimensional Yang-Mills.}
This is the key conjecture of the present paper which allows us to compute exactly the expectation value
of circular BPS 't~Hooft operator without reference to S-duality.  We do this computation in the next sections 
using the very well known partition function of 2d Yang-Mills on $S^2$ and we find explicitly precise agreement with S-duality predictions.

\subsubsection*{Elementary evaluation of $S_{YM}+S_{boundary}$ on classical solution}

On $D^3 \times_{\tilde w} S^1$ (see the metric (\ref{eq:metric-after-rescaling})) the SYM action \cite{Pestun:2009nn} evaluated on the classical solution (\ref{eq:tHooft-loop-singularity}) gives the integral 
\begin{multline}
  \label{eq:YM-action}
  S_{YM} = -\frac {1} {\gym^2} \int_{M_{\ve}} \tr \sqrt{g} (\frac 1 2 F_{\mu \nu} F^{\mu \nu} + D_{\mu} \Phi_A D^{\mu} \Phi^A + \frac {R}{6} \Phi_A \Phi^A) = \\
=\frac {1} {\gym^2} (2\pi) (4 \pi) \frac {\vec{m}^2} {4}  \int_{\ve}^{1} dx \, x^2 (\frac {1}{x^4} + \frac{1}{x^4} + \frac {2}{1 - x^2} \frac {1}{x^2})\frac 1 2 (1-x^2) = \frac {2 \pi^2 \vec{m}^2 }{\gym^2}(\frac 1 {\ve} - 1)\,,
\end{multline}
where we have regularized the integral by cutting out the region $x_i x_i < \epsilon^2$. The boundary term (\ref{eq:boundary-action}) evaluated on the resulting 3d boundary gives  
\begin{equation}
  S_{boundary} = - \frac{ 2 \pi^2 \vec{m}^2 } {\gym^2} \frac {1} {\ve},
\end{equation}
so the total action is
\begin{equation}
  \label{eq:total-elementary}
  S_{YM}(T_{\vec{m}}) + S_{boundary}(T_{\vec{m}}) =  - \frac {2 \pi^2 \vec{m}^2 }{\gym^2}\,.
\end{equation}
Notice that we get exactly the same result as for the 1/2 BPS circular loop on $\mathbb{R}^4$, see eq. (\ref{eq:result}), as expected by conformal invariance.

%\subsection{Localization in 2d YM}

\section{BPS 't~Hooft loops from 2d Yang-Mills unstable instantons\label{tHooft-2d}}
Let us consider 2d Yang-Mills theory on $S^2$ with gauge group $U(N)$. In our conventions, the action reads 
\begin{equation}
S=-\frac{1}{2 g^2} \int_{S^2} d^2x \sqrt{g} \tr F^2\,.
\label{2d-action}
\end{equation} 
It can be shown \cite{Witten:1992xu,Cordes:1994sd,Blau:1991mp,Blau:1993hj} that  two-dimensional Yang-Mills theory localizes on the classical configurations solving $D * F = 0$, called (unstable) instantons.  For $\U(N)$ gauge group each such configuration on $S^2$ is labeled by $N$ integers $m_1, \dots, m_N$. In the standard polar coordinates, the explicit instanton solutions may be written as the diagonal matrix
\begin{equation}
F_{inst} =  \frac{1}{2} \sin\theta d\theta \wedge d\phi \, T_{\vec{m}}\,.   
\label{instanton-sol}
\end{equation}

The exact partition function of  2d YM on $S^2$ has a representation as a sum over such configurations
\begin{equation}
Z^{YM_2}_{S^2}(g) = \sum_{m_i=-\infty}^{\infty} Z (g; m_1, \ldots, m_N).
\label{Z-inst-repr}
\end{equation} 
Each instanton configuration contributes with the usual classical weight $Z_{class}= e^{-S_{inst}}$ multiplied by  the  factor $Z_{quant}$ accounting for the quantum fluctuations \cite{MR674406,MR685019,MR1094734,MR721448}
\begin{equation}
  \label{eq:classical-quantum}
  Z(g; m_1,\ldots, m_N) = \exp(-S_{inst}(g; m_i)) Z_{quant}(g; m_1, \ldots, m_N),
\end{equation}
where 
\begin{equation}
S_{inst}(g;m_i) = \frac{4 \pi^2}{g^2A} \sum_{i=1}^N m_i^2,
\label{Sinst-class}
\end{equation}
is the classical action (\ref{2d-action}) evaluated on the instanton solution (\ref{instanton-sol}), and $A$ is the area of $S^2$. 
This agrees with our classical 4d computation (\ref{eq:total-elementary}) as supposed under the relation \cite{Drukker:2007yx,Drukker:2007dw,Pestun:2009nn}
\begin{equation}
  \label{eq:4d-2d}
  g^2 = - \frac{ 2 \gym^2 }{A}.
\end{equation}

The localization arguments discussed in the previous section lead us to propose that the exact expectation value of
 the 1/2 BPS circular 't~Hooft loop in representation $^L R =(m_1,\ldots,m_N)$ in ${\cal N}=4$ SYM with gauge group $G$
 can be computed from
 the partition function of 2d YM with gauge group $G$ around an unstable instanton labeled by $^L R$
\begin{equation}
\langle T_{{^L}R} ({\cal C}) \rangle  \leftrightarrow \frac{Z(g;m_1,\ldots,m_N)}{Z(g;0,\ldots,0)}\,,
\label{thooftVSinst}
\end{equation}
where the normalization by the 0-instanton partition function is such that the 't~Hooft loop in trivial representation has unit expectation value. Actually, because of the phenomenon known as ``monopole bubbling'' \cite{Kapustin:2006pk}, we expect that the ``naive'' 't~Hooft loop
 corresponding to a single unstable instanton in 2d YM according to (\ref{thooftVSinst}) will give the full exact result only for the case of the rank $k$ antisymmetric representation $R=A_k=(\stackrel{k}{\overbrace{1,\ldots,1}},0,\ldots,0)$ (including the fundamental as a special case). This is the representation with
 smallest $\sum_i m_i^2$ for fixed $\sum_i m_i$, and
 cannot be screened to give rise to subleading saddle points. For this reason, we will specialize to this choice of representation in the following, and leave the study of more general representations for future work.

In the localization context, the quantum factor $Z_{quant}(g; m_1, \ldots, m_N)$ usually has cohomological interpretation \cite{Beasley:2005vf}, and it can be exactly computed by the perturbation theory in the coupling constant $g$. 
The perturbative series actually terminates at finite order, so $Z_{quant}(g; m_1, \ldots, m_N)$ turns out to be a polynomial of finite degree in $g$ 
\cite{Witten:1992xu,Blau:1993hj,Blau:1991mp,Cordes:1994sd}.
%///I wonder if this polynomial (for example we know it is  Laguerre polynomial in the basic case $(1,0,\dots, 0)$)
%has been actually computed in the literature from the cohomology of the moduli space of unstable instantons. It would 
%be good to see this computation and we need reference. V.P.
However, perhaps a simpler way to obtain the instanton representation (\ref{Z-inst-repr}) is to start from the well known expression \cite{Migdal:1975zf,Rusakov:1990rs} of the exact partition function of 2d YM as a sum over irreducible representations of the gauge group and perform a certain Poisson resummation (see e.g. \cite{Minahan:1993tp,Caselle:1993gc,Gross:1994mr,Bassetto:1998sr}). In the following we briefly review this approach, following \cite{Bassetto:1998sr}. 

The exact partition function of 2d YM on $S^2$ is given by \cite{Migdal:1975zf,Rusakov:1990rs} 
\begin{equation}
Z_{S^2}^{YM_2}=\sum_R d_R^2 e^{-\frac{g^2 A}{4} C_2(R)}\,,
\label{exactZ}
\end{equation}
where $d_R$ is the dimension of the representation and $C_2(R)$ is the quadratic Casimir. 
Irreducible representations of $\U(N)$ are labeled in the standard way by Young diagrams $\vec{\lambda} = \lambda_{1} \geq \lambda_{2} \geq \dots \geq \lambda_{N}$ 
where $\lambda_{k}$ denote the lengths of rows. The character $\chi_{\vec{\lambda}}(\theta)$, which is defined
as trace in representation $\vec{\lambda}$ of the group element $\diag(z_i) \in G$, where $ z_i = e^{i \theta_i}$, is given 
by the Schur polynomial of $z_i$
\begin{equation}
  \label{eq:character-SU-N}
  \chi_{\vec{\lambda}}(e^{i\theta}) =  \frac { \det_{ij} e^{i\theta_i l_j }} { \det_{ij} e^{i\theta_i (N-j) }} , \quad l_j = \lambda_j + N - j, \quad i,j =1, \dots, N\,.
\end{equation}
Equivalently, irreducible representations of $\U(N)$ are labeled by strictly decreasing $N$-tuples of integers $\infty > l_1 > l_2 > \dots > l_N > - \infty$, 
with the character being given by the same formula (\ref{eq:character-SU-N}), and the dimension 
computed as
\begin{equation}
  \label{eq:dimension-UN-rep}
  d_{\lambda} = \frac { \Delta (l_1,\ldots,l_N)} { \Delta (N,\ldots,1) } =  \frac { \prod_{i < j} (\lambda_i - \lambda_j + j - i )} { \prod_{i < j} (j - i) },
\end{equation}
where $\Delta$ denotes the Vandermonde determinant 
\begin{equation}
\Delta(l_1,\ldots,l_N)=\prod_{i<j=1}^{N} (l_i-l_j)\,.
\end{equation}
The quadratic casimir $C_2(R)$ for $\U(N)$ is  
\begin{equation}
C_2(R)=-\frac{N}{12}(N^2-1) +\sum_{i=1}^{N} (l_i-\frac{N-1}{2})^2.
\end{equation}
Then (\ref{exactZ}) can be written explicitly as 
\begin{equation}
Z_{S^2}^{YM_2}=\frac{1}{N!}  c(N,g) \sum_{l_i=-\infty}^{\infty} \Delta^2(l_1,\ldots,l_N) e^{-\frac{g^2 A}{4} \sum_{i=1}^N (l_i-\frac{N-1}{2})^2},
\label{ZS2-msum}
\end{equation}
where  using antisymmetry of $\Delta$ with with respect to permutations of $l_i$ we extended the range of summation 
to arbitrary $N$-tuples of $l_{i} \in \BZ$, and
 by $c(N,g)$ we denoted the trivial factor 
\begin{equation}
  \label{eq:cNg}
  c(N,g) =  \frac{1} {( \prod_{i=1}^{N-1} i!)^2} e^{ \frac {g^2  A N(N^2-1)}{48}}.
\end{equation}

%where the integers $(m_1,\ldots,m_N)$ are related to the Young diagram for the representation $R$, and we have 
%used the expression for the quadratic Casimir

%/// I am sure that there must be minus sign in front of the 1/12 term. For Young diagram $d_1 \geq d_2 \dots \geq d_N$ labelling representation 
%in the standard way ( where we take $n$ rows of length $d_k$ and we symmetrize over rows and antisymmetrize over columns), 
%the integers $m_k = d_k + N - k$, and the $m_k$ strictly decrease. The fundamental rep is $d_1 = 0 , d_2 = d_3 = \dots = 0$. You can check 
%that for fundamental rep we should get $C_2 = N$ which is standard thing. 

We now perform a Poisson resummation of (\ref{ZS2-msum}), using the formula
\begin{equation}
\sum_{l_i=-\infty}^{\infty} f(l_1,\ldots,l_N) = \sum_{m_i=-\infty}^{\infty} \int_{-\infty}^{\infty} dz_1\ldots dz_N e^{2 \pi i \sum_{i=1}^N m_i z_i} f(z_1,\ldots,z_N)\,.
\label{Poisson}
\end{equation}
Therefore we have, after a simple shift in the $z$-variable
\begin{equation}
Z_{S^2}^{YM_2} = c(N,g) \frac{1}{N!} \sum_{m_i=-\infty}^{\infty} 
e^{i \pi (N-1) \sum_i m_i} \int d^N z \prod_{i <j} (z_i-z_j)^2 e^{-\frac{g^2 A}{4} \sum_{i=1}^N z_i^2} e^{2\pi i \sum_{i=1}^N m_i z_i}\,,
\label{Z-resummed}
\end{equation}

%/// This phase $\exp (i \pi (N-1)\sum_{n_i})$ is not uniquely fixed and can be removed
%if we want. Actually, to 2d YM action we can always add the $\theta$-term 
%$S_{top} = \frac{  \theta}{2 \pi} \int_{\Sigma} \tr  F$ which couples to the first Chern class of the gauge bundle on $\Sigma$ (we have $\frac{i}{2\pi} \in%t_{\Sigma} \tr F  = \sum_{i} m_i $).

Each term in the sum over the $m_i$'s is now physically interpreted as the contribution of an unstable instanton (\ref{instanton-sol}) with quantum numbers $(m_1,\ldots,m_N)$. To make the interpretation more transparent, one can notice that after elementary manipulations we can indeed rewrite (\ref{Z-resummed}) in the form (\ref{Z-inst-repr})-(\ref{eq:classical-quantum})
\begin{equation}
Z_{S^2}^{YM_2} = \sum_{m_i=-\infty}^{\infty} e^{-\frac{4 \pi^2}{g^2A} \sum_{i=1}^N m_i^2} Z_{quant}(g;m_1,\ldots,m_N)\,,
\end{equation}
where 
\begin{equation}
Z_{quant}(g;m_1,\ldots,m_N) = \frac{1}{N!} c(N,g) e^{i \pi (N-1) \sum_i m_i} \int d^N z \, \Delta^2\left(\vec{z}+\frac{4\pi i \vec{ m} }{g^2 A}\right) e^{-\frac{g^2 A}{4} \sum_{i=1}^N z_i^2} 
\end{equation}
corresponds to quantum fluctuations around the unstable instanton.

We now show that our proposal (\ref{thooftVSinst}) is precisely consistent with the S-duality symmetry of ${\cal N}=4$ SYM exchanging magnetic and electric loops. Recall that the expectation value of a circular 1/2 BPS Wilson loop in representation $^L R$ in  ${\cal N}=4$ SYM with gauge group $^L G$ and coupling 
constant $^L \gym$ is computed exactly by a matrix integral over the Lie algebra ${^L}\mathfrak{g}$ of ${^L}G$ \cite{Pestun:2007rz}. In the case of $^L G=G=U(N)$, this is the familiar Gaussian Hermitian matrix model \cite{Erickson:2000af,Drukker:2000rr}
\begin{equation}
\langle W_{^L R} ({\mathcal C}) \rangle = \frac{1}{{\cal Z}( ^L \gym^2 )} \int DX \, e^{-\frac{2}{ ^L \gym^2}\tr X^2} \frac{1}{d_{^L R}} \tr_{^L R} e^X\,, 
\label{Wilson-mm}
\end{equation}
where $g_{4d}$ is the SYM coupling constant, and ${\cal Z}({^L}g^2_{4d})$ is the matrix model partition function. According to S-duality, in the $U(N)$ theory the Wilson loop at coupling $^L \gym$ is mapped to the 't~Hooft loop at the dual coupling ${\gym^2 }=16\pi^2/{^L \gym^2}$ 
\begin{equation}
\langle W_{^L R}({\mathcal C}) \rangle_{^L \gym} = \langle T_{^L R} ({\mathcal C}) \rangle_{\gym}\,.
\end{equation}

The relation of the unstable instanton partition function to the Gaussian matrix model can be quickly recognized by looking at eq. (\ref{Z-resummed}). Making a simple change of variables $2\pi i z_i = x_i$ and plugging in the map $g^2 A= -2g^2_{4d}$ between 2d and 4d couplings \cite{Drukker:2007yx,Drukker:2007qr}, we have (dropping overall constants which do not depend on the $m_i$'s)
\begin{equation}
Z(g;m_1,\ldots,m_N)\simeq  
e^{i\pi(N-1) \sum_i m_i} \int d^N x \prod_{i <j} (x_i-x_j)^2 e^{-\frac{g^2_{4d}}{8\pi^2} \sum_{i=1}^N x_i^2} e^{\sum_i m_i x_i}\,.
\end{equation}
The integration over the $x_i$ variables is clearly equivalent to the integration over eigenvalues in a Gaussian Hermitian matrix 
model with potential $V(X)=\frac{g^2_{4d}}{8\pi^2} \tr X^2$. Moreover, specializing to
 the rank $k$ antisymmetric representation, one can see that the insertion of $e^{\sum_i m_i x_i}=e^{\sum_{i=1}^k x_i}$ is
 equivalent to the insertion of the character $\frac{1}{d_{A_k}} \tr_{A_k} e^X$ in the matrix model, since in the eigenvalue basis
\begin{equation}
\frac{1}{d_{A_k}} \tr_{A_k} e^X =\frac{(N-k)!k!}{N!} \sum_{i_1 < i_2 <\cdots <i_k} e^{x_{i_1}+x_{i_2}+\ldots +x_{i_k}}\,.
\end{equation}
But the integrand is symmetric under permutations of the $x_i$'s, hence we can just take one term in the sum above multiplied by $d_{A_k}$, and we exactly end up with the insertion of $e^{\sum_{i=1}^k x_i}$ in the eigenvalue integral. Putting everything together, the identification (\ref{thooftVSinst}) implies the exact prediction for the 't~Hooft loop expectation value 
\begin{equation}
\langle T_{A_k}({\mathcal C}) \rangle =  \frac{(-1)^{k(N-1)}}{{\cal Z}({^Lg^2_{4d}})} \int DX \, e^{-\frac{2}{{^Lg^2_{4d}}}\tr X^2} \frac{1}{d_{A_k}} \tr_{A_k} e^X\,, \qquad {^Lg^2_{4d}} = \frac{16 \pi^2}{g^2_{4d}}\,.
\end{equation}
This is precisely equal to the Wilson loop expectation value (\ref{Wilson-mm}) in the same representation and with dual coupling constant, up to the overall sign. Fixing this sign requires a more careful study of the one-loop determinant for fluctuations around the supersymmetric configurations. Notice that the correct form of the S-dual coupling ${^Lg^2_{4d}}$, including the numerical factor, is correctly predicted by the 2d YM unstable instanton partition function.

As an example, in the case of the fundamental representation we can compute the integral over eigenvalues explicitly by using orthogonal polynomials (see e.g. \cite{Drukker:2000rr}), and we get the exact result 
\begin{equation}
\langle T_{F}({\mathcal C}) \rangle_{\gym} = \frac{(-1)^{(N-1)}}{N} L_{N-1}^1 \left(-\frac{4\pi^2}{g^2_{4d}}\right)  e^{\frac{2\pi^2}{g^2_{4d}}}\,, 
\end{equation}
where $L_{N-1}^1(x)$ is a Laguerre polynomial\footnote{The Laguerre polynomials can be defined by $L_n^k(x)=\frac{x^{-k}e^x}{n!}\frac{d^n}{dx^n}\left(e^{-x}x^{n+k}\right)$. We also denote $L_n^0(x)\equiv L_n(x)$.}. From the point of view of the 2d YM instanton partition function, the exponential factor corresponds to the classical action (\ref{Sinst-class}), while the Laguerre polynomial comes from the quantum corrections around the instanton.

\section{Wilson-'t~Hooft correlator\label{Wilson-tHooft}}
According to our conjecture, we can also compute correlation functions of the 1/2 BPS 't~Hooft loop with any number of 1/8 BPS Wilson loops inserted on the $S^2$ linked to the 't~Hooft loop. This is simply done in the 2d theory by calculating the Wilson loop correlation functions around a fixed unstable instanton. As an example, we compute here the correlator of the 't~Hooft loop and a Wilson loop in the case in which both operators are in the fundamental representation of $U(N)$. We leave the study of more general representations (and gauge groups) to future study.

Let us start from the exact expression for the expectation value of a Wilson loop in the 2d Yang-Mills theory on $S^2$ \cite{Migdal:1975zf,Rusakov:1990rs}
\begin{equation}
\langle W_R(A_1,A_2)\rangle^{YM_2}_{S^2}=\int dU \sum_{R_1,R_2} d_{R_1} d_{R_2} \chi_{R_1}(U) \bar\chi_{R_2}(U) e^{-\frac{g^2A_1}{4} C_2(R_1)-\frac{g^2A_2}{4}C_2(R_2)} \frac{1}{d_R}\tr_R U\,, 
\end{equation}
where the integral is taken over the $U(N)$ group manifold and $\chi_{R_i}(U)$ denotes the character of $U$ in representation $R_i$. Here $A_1,A_2$ are the areas of the two regions singled out by the loop on $S^2$.
 
Specializing to the case of Wilson loop in the fundamental, after performing the integration over $U(N)$, this can be written explicitly as \cite{Bassetto:1998sr}
\begin{equation}
\begin{aligned}
\langle W_F(A_1,A_2)\rangle^{YM_2}_{S^2}&=c(N,g) \frac{1}{N!}  \sum_{k=1}^N \sum_{l_i=-\infty}^{\infty} \Delta(l_i)\Delta(l_i+\delta_{ik}) \times \\
&\times e^{-\frac{g^2 A_1}{4} \sum_{i=1}^N (l_i-\frac{N-1}{2})^2-\frac{g^2 A_2}{4} \sum_{i=1}^N (l_i-\frac{N-1}{2}+\delta_{ik})^2}\,.
\end{aligned}
\end{equation} 
To obtain the instanton expansion of this result, we again perform a Poisson resummation using (\ref{Poisson}), and we get
\begin{equation}
\begin{aligned}
&\langle W_F(A_1,A_2)\rangle^{YM_2}_{S^2} = \sum_{m_i=-\infty}^{\infty} \langle W_F(A_1,A_2) \rangle_{(\vec{m})}\,,\\
&\langle W_F(A_1,A_2) \rangle_{(\vec{m})}=c(N,g) \frac{1}{N!} e^{i \pi (N-1)\sum_i m_i}  \times \\ 
& \qquad \qquad \times \sum_{k=1}^N \int d^N z\Delta(z_i)\Delta(z_i+\delta_{ik}) e^{2\pi i \sum_i m_i z_i} e^{-\frac{g^2 A_1}{4} \sum_{i=1}^N z_i^2-\frac{g^2 A_2}{4} \sum_{i=1}^N (z_i+\delta_{ik})^2}\,.
\end{aligned}
\label{W-inst}
\end{equation}
In this formula, $\langle W_F(A_1,A_2) \rangle_{(\vec{m})}$ corresponds to the Wilson loop average around an unstable instanton with quantum numbers $\vec{m}=(m_1,\ldots,m_N)$, and hence, according to our conjecture, it gives the 4d correlator between the 1/8 BPS Wilson loop and the 1/2 BPS 't~Hooft loop labeled by $\vec{m}$\footnote{Modulo the issue of monopole bubbling discussed above, i.e. we expect the naive equivalence to be exact only for 't~Hooft loop in the ``minuscule'' representations.}.

As an example, we now evaluate explicitly the integral in (\ref{W-inst}) in the case of the instanton/'t~Hooft
 loop in the fundamental $\vec{m}={(1,0,\ldots,0)}$. To simplify the equations, we
 will also restrict in the following to the 1/2 BPS great circle with $A_1=A_2=A/2$, but the generalization to arbitrary area is straightforward.

Due to the symmetries of the integrand, the sum over $k$ in (\ref{W-inst}) can be reduced to two terms: the one with $k=1$ and the one e.g. with $k=2$ counted $N-1$ times. The integrals can be performed explicitly using the standard trick of rewriting the Vandermonde determinants in terms of orthogonal polynomials (in the present case it is convenient to use Hermite polynomials, due to the Gaussian integration measure). In evaluating the integrals, the following identity involving Hermite polynomials\footnote{The Hermite polynomial are given by $H_k(x)=(-1)^k e^{x^2}\frac{d^k}{dx^k}e^{-x^2}$. The formula (\ref{H-ide})  can be proven for example by using the identity $H_k(x+a)=(H+2a)^k$, where it is understood that $H^k\equiv H_k(x)$.} turns out to be useful
\begin{equation}
\int_{-\infty}^{\infty} dx e^{-x^2} H_k(x+a) H_l(x+b) = 2^k \sqrt{\pi} k! (2b)^{l-k} L_k^{l-k}(-2ab)\,,\quad k \le l\,,
\label{H-ide}
\end{equation}
where $L_k^{l-k}(-2ab)$ is a Laguerre polynomial.

After normalizing $\langle W_F(A/2,A/2) \rangle_{(1,0,\ldots,0)}$ by the 0-instanton partition function, we finally obtain our prediction for the exact correlator of the 1/2 BPS 't~Hooft loop and the 1/2 BPS Wilson loop in the 4d ${\cal N}=4$ SYM theory
%Our final result is 
%\begin{equation}
%\begin{aligned}
%&\langle T_F({\mathcal C}) W_F ({\mathcal C}')\rangle = \frac{(-1)^{N-1}e^{g^2_{4d}\frac{A_1 %A_2}{2A^2}+\frac{2 \pi^2}{g^2_{4d}}}}{N^2} \Bigg{\{} e^{-\frac{2 \pi i A_2}{A}} %L_{N-1}^1\left(-g^2_{4d}\frac{A_1 A_2}{A^2}-\frac{4 \pi^2}{g^2_{4d}} + 2\pi i n \frac{A_2-A_1}{A}\right)\\
%& + L_{N-1}^1\left(-g^2_{4d}\frac{A_1 A_2}{A^2}\right)L_{N-1}^1\left(-\frac{4 \pi^2}{g^2_{4d}}\right)
%-\sum_{j=1}^N L_{j-1}\left(-g^2_{4d}\frac{A_1 A_2}{A^2}\right)L_{j-1}\left(-\frac{4 %\pi^2}{g^2_{4d}}\right)\\
%& -\sum_{j_1 < j_2 =1}^N \frac{(j_1-1)!}{(j_2-1)!} \left[\left(\frac{2\pi i %A_1}{A}\right)^{j_2-j_1}+\left(\frac{-2\pi i A_2}{A}\right)^{j_2-j_1} \right] %L_{j_1-1}^{j_2-j_1}\left(-g^2_{4d}\frac{A_1 A_2}{A^2}\right)L_{j_1-1}^{j_2-j_1}\left(-\frac{4 %\pi^2}{g^2_{4d}}\right)
%\Bigg{\}}
%\end{aligned}
%\end{equation}
\begin{equation}
\begin{aligned}
&\langle T_F({\mathcal C}) W_F ({\mathcal C}')\rangle = (-1)^{N-1} \frac{e^{\frac{g^2_{4d}}{8}+\frac{2 \pi^2}{g^2_{4d}}}}{N^2} \Bigg{\{} -L_{N-1}^1\left(-\frac{g^2_{4d}}{4}-\frac{4 \pi^2}{g^2_{4d}}\right)\\
& + L_{N-1}^1\left(-\frac{g^2_{4d}}{4}\right)L_{N-1}^1\left(-\frac{4 \pi^2}{g^2_{4d}}\right)
-\sum_{j=1}^N L_{j-1}\left(-\frac{g^2_{4d}}{4}\right)L_{j-1}\left(-\frac{4 \pi^2}{g^2_{4d}}\right)\\
& -\sum_{j_1 < j_2 =1}^N \frac{(j_1-1)!}{(j_2-1)!} \left[(i \pi)^{j_2-j_1}+(-i \pi)^{j_2-j_1}\right] L_{j_1-1}^{j_2-j_1}\left(-\frac{g^2_{4d}}{4}\right)L_{j_1-1}^{j_2-j_1}\left(-\frac{4 \pi^2}{g^2_{4d}}\right)
\Bigg{\}}\,,
\end{aligned}
\end{equation}
where ${\cal C}'$ denotes the circle at the equator of $S^2$. Notice that the correlation function is invariant under the $S$-dual replacement $g^2_{4d} \rightarrow {^Lg^2_{4d}}=16\pi^2/g^2_{4d}$ 
\begin{equation}
\langle T_F({\mathcal C}) W_F ({\mathcal C}')\rangle_{g^2_{4d}} = \langle T_F({\mathcal C}) W_F ({\mathcal C}')\rangle_{^Lg^2_{4d}} \,.
\end{equation}
This is precisely as expected, since $S$-duality exchanges the roles of Wilson and 't~Hooft loop in the correlation function\footnote{Of course, $S$-duality does not map $T_F({\mathcal C})$ and $W_F ({\mathcal C}')$ to each other, but it maps the 't~Hooft loop to a Wilson loop on the same circle and vice-versa. But the correlation function is insensitive to whether we put the Wilson/'t~Hooft loop on ${\cal C}$/${\cal C}'$.}.

For future reference, we also quote here the small coupling expansion of the result 
\begin{equation}
\langle T_F({\mathcal C}) W_F ({\mathcal C}')\rangle = (-1)^{N-1}e^{\frac{2 \pi^2}{g^2_{4d}}} \left(\frac{4\pi^2}{g_{4d}^2}\right)^{N-1}  \frac{(N-2)}{N^2(N-1)!}\left[1+\frac{g^2_{4d} N}{8 \pi^2}(\pi^2+2(N-1))+\ldots \right]\,.
\end{equation}
%It would be interesting to check this by a direct perturbative calculation in the 4d gauge theory.
We notice that the Wilson loop in the 't~Hooft loop background normalized by the expectation value of the 't~Hooft loop has perturbative 
expansion 
\begin{equation}
  \label{eq:WT-over-T-perturbative}
 \frac{  \langle T_F({\mathcal C}) W_F ({\mathcal C}')\rangle } {  \langle T_F({\mathcal C}) \rangle} = 
\frac{ N -2} {N} + (N-2) g_{4d}^2  + \dots. 
\end{equation}
It it easy to check that (\ref{eq:WT-over-T-perturbative}) agrees with the $\CalN=4$ SYM perturbation theory for $W$ in the background 
of $T$. The first term in (\ref{eq:WT-over-T-perturbative}) is obtained as a classical value of $W(\mathcal{C}')$ in the background of $T(\mathcal{C})$,
and the second term in (\ref{eq:WT-over-T-perturbative}) comes from the one-ladder diagram. The background of 't~Hooft loop breaks the $\U(N)$ to 
$\U(N-1) \times \U(1)$. The diagonal $\U(1) \times \U(1)$ and $\U(N-1) \times \U(N-1)$ blocks for the 
propagators of the relevant gauge and scalar fields in this background are unchanged and contribute respectively 
as $-g_{4d}^2/8$ and $(N-1)g_{4d}^2/8$ to
 the one-ladder diagram.  The correlators of the anti-diagonal blocks $\U(1)\times U(N-1)$ do not contribute to the expectation 
value of $W({\cal C}')$, so we get (\ref{eq:WT-over-T-perturbative}).

\bibliography{bsample}

\def\cprime{$'$}
\providecommand{\href}[2]{#2}\begingroup\raggedright\begin{thebibliography}{10}

\bibitem{Montonen:1977sn}
C.~Montonen and D.~I. Olive, ``Magnetic monopoles as gauge particles?,'' {\em
  Phys. Lett.} {\bf B72} (1977) 117.

\bibitem{Goddard:1976qe}
P.~Goddard, J.~Nuyts, and D.~I. Olive, ``{Gauge Theories and Magnetic
  Charge},'' {\em Nucl. Phys.} {\bf B125} (1977) 1.

\bibitem{Witten:1978mh}
E.~Witten and D.~I. Olive, ``{Supersymmetry Algebras That Include Topological
  Charges},'' {\em Phys. Lett.} {\bf B78} (1978) 97.

\bibitem{Osborn:1979tq}
H.~Osborn, ``{Topological Charges for N=4 Supersymmetric Gauge Theories and
  Monopoles of Spin 1},'' {\em Phys. Lett.} {\bf B83} (1979) 321.

\bibitem{Kapustin:2006pk}
A.~Kapustin and E.~Witten, ``{Electric-magnetic duality and the geometric
  Langlands program},'' \href{http://xxx.lanl.gov/abs/hep-th/0604151}{{\tt
  hep-th/0604151}}.

\bibitem{Argyres:2006qr}
P.~C. Argyres, A.~Kapustin, and N.~Seiberg, ``{On S-duality for
  non-simply-laced gauge groups},'' {\em JHEP} {\bf 06} (2006) 043,
  \href{http://xxx.lanl.gov/abs/hep-th/0603048}{{\tt hep-th/0603048}}.

\bibitem{Witten:2009mh}
E.~Witten, ``{Geometric Langlands And The Equations Of Nahm And Bogomolny},''
  \href{http://xxx.lanl.gov/abs/0905.4795}{{\tt 0905.4795}}.

\bibitem{Kapustin:2005py}
A.~Kapustin, ``{W}ilson-'t {H}ooft operators in four-dimensional gauge theories
  and {S}-duality,'' {\em Phys. Rev.} {\bf D74} (2006) 025005,
  \href{http://xxx.lanl.gov/abs/hep-th/0501015}{{\tt hep-th/0501015}}.

\bibitem{Erickson:2000af}
J.~K. Erickson, G.~W. Semenoff, and K.~Zarembo, ``{W}ilson loops in {{N =}4}
  supersymmetric {Y}ang-{M}ills theory,'' {\em Nucl. Phys.} {\bf B582} (2000)
  155--175, \href{http://xxx.lanl.gov/abs/hep-th/0003055}{{\tt
  hep-th/0003055}}.

\bibitem{Drukker:2000rr}
N.~Drukker and D.~J. Gross, ``An exact prediction of {{N =}4} {SUSYM} theory
  for string theory,'' {\em J. Math. Phys.} {\bf 42} (2001) 2896--2914,
  \href{http://xxx.lanl.gov/abs/hep-th/0010274}{{\tt hep-th/0010274}}.

\bibitem{Pestun:2007rz}
V.~Pestun, ``{Localization of gauge theory on a four-sphere and supersymmetric
  Wilson loops},'' \href{http://xxx.lanl.gov/abs/0712.2824}{{\tt 0712.2824}}.

\bibitem{Drukker:2007dw}
N.~Drukker, S.~Giombi, R.~Ricci, and D.~Trancanelli, ``More supersymmetric
  {W}ilson loops,'' \href{http://xxx.lanl.gov/abs/arXiv:0704.2237
  [hep-th]}{{\tt arXiv:0704.2237 [hep-th]}}.

\bibitem{Drukker:2007qr}
N.~Drukker, S.~Giombi, R.~Ricci, and D.~Trancanelli, ``Supersymmetric {W}ilson
  loops on {$S^3$},'' \href{http://xxx.lanl.gov/abs/arXiv:0711.3226
  [hep-th]}{{\tt arXiv:0711.3226 [hep-th]}}.

\bibitem{Drukker:2007yx}
N.~Drukker, S.~Giombi, R.~Ricci, and D.~Trancanelli, ``{W}ilson loops: From
  four-dimensional {SYM} to two-dimensional {YM},''
  \href{http://xxx.lanl.gov/abs/arXiv:0707.2699 [hep-th]}{{\tt arXiv:0707.2699
  [hep-th]}}.

\bibitem{Bassetto:1998sr}
A.~Bassetto and L.~Griguolo, ``{Two-dimensional {QCD}, instanton contributions
  and the perturbative Wu-Mandelstam-Leibbrandt prescription},'' {\em Phys.
  Lett.} {\bf B443} (1998) 325--330,
  \href{http://xxx.lanl.gov/abs/hep-th/9806037}{{\tt hep-th/9806037}}.

\bibitem{Bassetto:1999dg}
A.~Bassetto, L.~Griguolo, and F.~Vian, ``{Instanton contributions to Wilson
  loops with general winding number in two dimensions and the spectral
  density},'' {\em Nucl. Phys.} {\bf B559} (1999) 563--590,
  \href{http://xxx.lanl.gov/abs/hep-th/9906125}{{\tt hep-th/9906125}}.

\bibitem{Young:2008ed}
D.~Young, ``{BPS Wilson Loops on $S^2$ at Higher Loops},'' {\em JHEP} (2008)
  077, \href{http://xxx.lanl.gov/abs/0804.4098}{{\tt 0804.4098}}.

\bibitem{Bassetto:2008yf}
A.~Bassetto, L.~Griguolo, F.~Pucci, and D.~Seminara, ``{Supersymmetric Wilson
  loops at two loops},'' {\em JHEP} {\bf 06} (2008) 083,
  \href{http://xxx.lanl.gov/abs/0804.3973}{{\tt 0804.3973}}.

\bibitem{Giombi:2009ms}
S.~Giombi, V.~Pestun, and R.~Ricci, ``{Notes on supersymmetric Wilson loops on
  a two-sphere},'' \href{http://xxx.lanl.gov/abs/0905.0665}{{\tt 0905.0665}}.

\bibitem{Bassetto:2009rt}
A.~Bassetto {\em et.~al.}, ``{Correlators of supersymmetric Wilson-loops,
  protected operators and matrix models in N=4 SYM},''
  \href{http://xxx.lanl.gov/abs/0905.1943}{{\tt 0905.1943}}.

\bibitem{Pestun:2009nn}
V.~Pestun, ``{Localization of the four-dimensional N=4 SYM to a two- sphere and
  1/8 BPS Wilson loops},'' \href{http://xxx.lanl.gov/abs/0906.0638}{{\tt
  0906.0638}}.

\bibitem{Moore:1997dj}
G.~W. Moore, N.~Nekrasov, and S.~Shatashvili, ``{Integrating over Higgs
  branches},'' {\em Commun. Math. Phys.} {\bf 209} (2000) 97--121,
  \href{http://xxx.lanl.gov/abs/hep-th/9712241}{{\tt hep-th/9712241}}.

\bibitem{Gerasimov:2006zt}
A.~A. Gerasimov and S.~L. Shatashvili, ``{Higgs bundles, gauge theories and
  quantum groups},'' {\em Commun. Math. Phys.} {\bf 277} (2008) 323--367,
  \href{http://xxx.lanl.gov/abs/hep-th/0609024}{{\tt hep-th/0609024}}.

\bibitem{Gerasimov:2007ap}
A.~A. Gerasimov and S.~L. Shatashvili, ``{Two-dimensional Gauge Theories and
  Quantum Integrable Systems},'' \href{http://xxx.lanl.gov/abs/0711.1472}{{\tt
  0711.1472}}.

\bibitem{Nekrasov:2009rc}
N.~A. Nekrasov and S.~L. Shatashvili, ``{Quantization of Integrable Systems and
  Four Dimensional Gauge Theories},''
  \href{http://xxx.lanl.gov/abs/0908.4052}{{\tt 0908.4052}}.

\bibitem{Giombi:2009ds}
S.~Giombi and V.~Pestun, ``{Correlators of local operators and 1/8 BPS Wilson
  loops on $S^2$ from 2d YM and matrix models},''
  \href{http://xxx.lanl.gov/abs/0906.1572}{{\tt 0906.1572}}.

\bibitem{Gomis:2009ir}
J.~Gomis, T.~Okuda, and D.~Trancanelli, ``{Quantum 't Hooft operators and
  S-duality in N=4 super Yang-Mills},''
  \href{http://xxx.lanl.gov/abs/0904.4486}{{\tt 0904.4486}}.

\bibitem{Cherkis:2007jm}
S.~A. Cherkis and B.~Durcan, ``{The 't Hooft-Polyakov Monopole in the Presence
  of an 't Hooft Operator},'' {\em Phys. Lett.} {\bf B671} (2009) 123--127,
  \href{http://xxx.lanl.gov/abs/0711.2318}{{\tt 0711.2318}}.

\bibitem{Kapustin:2007wm}
A.~Kapustin and N.~Saulina, ``{The algebra of Wilson-'t Hooft operators},''
  {\em Nucl. Phys.} {\bf B814} (2009) 327--365,
  \href{http://xxx.lanl.gov/abs/0710.2097}{{\tt 0710.2097}}.

\bibitem{Intriligator:1998ig}
K.~A. Intriligator, ``{Bonus symmetries of N = 4 super-Yang-Mills correlation
  functions via AdS duality},'' {\em Nucl. Phys.} {\bf B551} (1999) 575--600,
  \href{http://xxx.lanl.gov/abs/hep-th/9811047}{{\tt hep-th/9811047}}.

\bibitem{Intriligator:1999ff}
K.~A. Intriligator and W.~Skiba, ``{Bonus symmetry and the operator product
  expansion of N = 4 super-Yang-Mills},'' {\em Nucl. Phys.} {\bf B559} (1999)
  165--183, \href{http://xxx.lanl.gov/abs/hep-th/9905020}{{\tt
  hep-th/9905020}}.

\bibitem{Gomis:2009xg}
J.~Gomis and T.~Okuda, ``{S-duality, 't Hooft operators and the operator
  product expansion},'' \href{http://xxx.lanl.gov/abs/0906.3011}{{\tt
  0906.3011}}.

\bibitem{Gukov:2006jk}
S.~Gukov and E.~Witten, ``{Gauge theory, ramification, and the geometric
  langlands program},'' \href{http://xxx.lanl.gov/abs/hep-th/0612073}{{\tt
  hep-th/0612073}}.

\bibitem{Gomis:2007fi}
J.~Gomis and S.~Matsuura, ``{Bubbling surface operators and S-duality},'' {\em
  JHEP} {\bf 06} (2007) 025, \href{http://xxx.lanl.gov/abs/0704.1657}{{\tt
  0704.1657}}.

\bibitem{Drukker:2008wr}
N.~Drukker, J.~Gomis, and S.~Matsuura, ``{Probing N=4 SYM With Surface
  Operators},'' \href{http://xxx.lanl.gov/abs/0805.4199}{{\tt 0805.4199}}.

\bibitem{Alday:2009fs}
L.~F. Alday, D.~Gaiotto, S.~Gukov, Y.~Tachikawa, and H.~Verlinde, ``{Loop and
  surface operators in N=2 gauge theory and Liouville modular geometry},''
  \href{http://xxx.lanl.gov/abs/0909.0945}{{\tt 0909.0945}}.

\bibitem{Drukker:2009id}
N.~Drukker, J.~Gomis, T.~Okuda, and J.~Teschner, ``{Gauge Theory Loop Operators
  and Liouville Theory},'' \href{http://xxx.lanl.gov/abs/0909.1105}{{\tt
  0909.1105}}.

\bibitem{Alday:2009aq}
L.~F. Alday, D.~Gaiotto, and Y.~Tachikawa, ``{Liouville Correlation Functions
  from Four-dimensional Gauge Theories},''
  \href{http://xxx.lanl.gov/abs/0906.3219}{{\tt 0906.3219}}.

\bibitem{Maldacena:1998im}
J.~M. Maldacena, ``{W}ilson loops in large {N} field theories,'' {\em Phys.
  Rev. Lett.} {\bf 80} (1998) 4859--4862,
  \href{http://xxx.lanl.gov/abs/hep-th/9803002}{{\tt hep-th/9803002}}.

\bibitem{Zarembo:2002an}
K.~Zarembo, ``Supersymmetric {W}ilson loops,'' {\em Nucl. Phys.} {\bf B643}
  (2002) 157--171, \href{http://xxx.lanl.gov/abs/hep-th/0205160}{{\tt
  hep-th/0205160}}.

\bibitem{Bogomolny:1975de}
E.~B. Bogomolny, ``{Stability of Classical Solutions},'' {\em Sov. J. Nucl.
  Phys.} {\bf 24} (1976) 449.

\bibitem{Witten:1992xu}
E.~Witten, ``Two-dimensional gauge theories revisited,'' {\em J. Geom. Phys.}
  {\bf 9} (1992) 303--368, \href{http://xxx.lanl.gov/abs/hep-th/9204083}{{\tt
  hep-th/9204083}}.

\bibitem{Cordes:1994sd}
S.~Cordes, G.~W. Moore, and S.~Ramgoolam, ``Large {N 2-D} {Y}ang-{M}ills theory
  and topological string theory,'' {\em Commun. Math. Phys.} {\bf 185} (1997)
  543--619, \href{http://xxx.lanl.gov/abs/hep-th/9402107}{{\tt
  hep-th/9402107}}.

\bibitem{Blau:1991mp}
M.~Blau and G.~Thompson, ``{Quantum Yang-Mills theory on arbitrary surfaces},''
  {\em Int. J. Mod. Phys.} {\bf A7} (1992) 3781--3806.

\bibitem{Blau:1993hj}
M.~Blau and G.~Thompson, ``{Lectures on 2-d gauge theories: Topological aspects
  and path integral techniques},''
  \href{http://xxx.lanl.gov/abs/hep-th/9310144}{{\tt hep-th/9310144}}.

\bibitem{MR674406}
J.~J. Duistermaat and G.~J. Heckman, ``On the variation in the cohomology of
  the symplectic form of the reduced phase space,'' {\em Invent. Math.} {\bf
  69} (1982), no.~2 259--268.

\bibitem{MR685019}
N.~Berline and M.~Vergne, ``Classes caract\'eristiques \'equivariantes.
  {F}ormule de localisation en cohomologie \'equivariante,'' {\em C. R. Acad.
  Sci. Paris S\'er. I Math.} {\bf 295} (1982), no.~9 539--541.

\bibitem{MR1094734}
M.~F. Atiyah and L.~Jeffrey, ``Topological {L}agrangians and cohomology,'' {\em
  J. Geom. Phys.} {\bf 7} (1990), no.~1 119--136.

\bibitem{MR721448}
M.~F. Atiyah and R.~Bott, ``The moment map and equivariant cohomology,'' {\em
  Topology} {\bf 23} (1984), no.~1 1--28.

\bibitem{Beasley:2005vf}
C.~Beasley and E.~Witten, ``{Non-abelian localization for Chern-Simons
  theory},'' {\em J. Diff. Geom.} {\bf 70} (2005) 183--323,
  \href{http://xxx.lanl.gov/abs/hep-th/0503126}{{\tt hep-th/0503126}}.

\bibitem{Migdal:1975zf}
A.~A. Migdal, ``{Gauge Transitions in Gauge and Spin Lattice Systems},'' {\em
  Sov. Phys. JETP} {\bf 42} (1975) 743.

\bibitem{Rusakov:1990rs}
B.~E. Rusakov, ``{Loop averages and partition functions in U(N) gauge theory on
  two-dimensional manifolds},'' {\em Mod. Phys. Lett.} {\bf A5} (1990)
  693--703.

\bibitem{Minahan:1993tp}
J.~A. Minahan and A.~P. Polychronakos, ``{Classical solutions for
  two-dimensional QCD on the sphere},'' {\em Nucl. Phys.} {\bf B422} (1994)
  172--194, \href{http://xxx.lanl.gov/abs/hep-th/9309119}{{\tt
  hep-th/9309119}}.

\bibitem{Caselle:1993gc}
M.~Caselle, A.~D'Adda, L.~Magnea, and S.~Panzeri, ``{Two-dimensional QCD is a
  one-dimensional Kazakov-Migdal model},'' {\em Nucl. Phys.} {\bf B416} (1994)
  751--770, \href{http://xxx.lanl.gov/abs/hep-th/9304015}{{\tt
  hep-th/9304015}}.

\bibitem{Gross:1994mr}
D.~J. Gross and A.~Matytsin, ``{Instanton induced large N phase transitions in
  two- dimensional and four-dimensional QCD},'' {\em Nucl. Phys.} {\bf B429}
  (1994) 50--74, \href{http://xxx.lanl.gov/abs/hep-th/9404004}{{\tt
  hep-th/9404004}}.

\end{thebibliography}\endgroup
\end{document}